\documentclass[final,12pt,authoryear,times]{elsarticle}
\usepackage[utf8]{inputenc}
\usepackage{amsmath}
\usepackage{mathrsfs}
\usepackage{graphicx}
\usepackage{epstopdf}
\usepackage{enumitem}
\usepackage{siunitx}
\usepackage{float}
\usepackage{mathtools}
\usepackage{xcolor}
\usepackage{subcaption}
\usepackage{algorithm}
\usepackage{algorithmic}
\usepackage[hidelinks]{hyperref}
\usepackage{textcomp} 
\usepackage{stmaryrd}
\newcommand{\comment}[1]{} 

 \usepackage[modulo]{lineno}

 \newcommand{\dd}{\mathrm{d}}

\usepackage{amssymb}
\usepackage{graphicx}
\usepackage{float}
\usepackage[english]{babel}
\usepackage[normalem]{ulem}
\usepackage{amsmath}
\usepackage{amsthm}
\usepackage{amsfonts}
\usepackage{enumerate}
\usepackage{bm}

\usepackage{physics}
\usepackage{lineno}

\journal{Journal of the Mechanics and Physics of Solids}

\begin{document}



\begin{frontmatter}

\title{A Unified Thermo-Chemo-Mechanical Framework for Bulk and Frontal Polymerization: Coupled Kinetics and Front Stability
}



\author[label1]{Xuanhe Li}
\author[label1,label2]{Tal Cohen\corref{cor1}}
\cortext[cor1]{Corresponding author: talco@mit.edu}

\address[label1]{Massachusetts Institute of Technology, Department of Mechanical Engineering, Cambridge, MA, 02139, USA}
\address[label2]{Massachusetts Institute of Technology, Department of Civil and Environmental Engineering, Cambridge, MA, 02139, USA}


\begin{abstract}
Polymerization is a fundamental chemical process that enables large-scale production of material components that are essential in nearly all modern industries. In transforming a monomer mixture into a cross-linked structure of polymer chains, polymerization introduces   changes in temperature and in material properties, such as density and stiffness, which can induce residual stress and warping via coupled mechanisms that are not yet fully understood. Depending on the specific conditions, polymerization can occur in the bulk of the material, with the reaction sustained by the continuous injection of energy via heat or light, or across a self-sustaining exothermic reaction front. The latter is commonly referred to as `frontal polymerization' and has attracted increasing interest as an energy-efficient and rapid curing method. However, the local nature of the polymerization reaction in these systems is accompanied by abrupt spatial variations  thus amplifying thermo-chemo-mechanical coupling effects. In this paper, we develop a generalized thermodynamically consistent framework that seamlessly captures both bulk and frontal polymerization. The formulation incorporates both the influence of local stress on reaction kinetics, and the evolution of the stress-free configuration during curing. The latter is developed in analogy to existing models for plasticity theory to  establish a kinetic relation for the evolution of irreversible deformations, akin to a flow rule.  In our analysis, we show how specific material properties govern the transition from bulk to frontal polymerization.  Using a narrow reaction zone approximation in a uniaxial setting, we  obtain analytical predictions for the propagation velocity, and examine the role of residual stresses during front propagation in comparison with experiments. A perturbation analysis reveals a stability condition for front propagation and enables the construction of  a phase diagram which illustrates  the  influence of heat loss and mechanical loading on transitions between regimes of stable and unstable  front propagation, and a regime where frontal polymerization is impossible. This stability criterion is shown to  be a generalization of the classical `Zeldovich number' that applies  the adiabatic, stress-free limit.  
Overall, this framework  both offers a tool to guide future precision polymerization based manufacturing methods that can  manage  residual stresses and heat, and reveals fundamental sensitivities of polymerization processes, which can inform future synthesis efforts.

\end{abstract}


\end{frontmatter}



\section{Introduction}

\noindent
Polymerization, also known as polymer curing, is a process in which a liquid monomer mixture is transformed into a polymer network. This transformation enables the fabrication of polymer components with a broad range of applications, spanning from small-scale biomedical devices \citep{ramakrishna2001biomedical} and protective coatings \citep{kausar2018polymer} to large-scale structural components such as wind turbine blades \citep{mishnaevsky2017materials} and aircraft wings \citep{Boeing}. Beyond traditional manufacturing, polymerization also serves as the foundational mechanism in additive manufacturing technologies such as 3D printing \citep{bagheri2019photopolymerization}. As industrial demands continue to evolve, there is a growing emphasis on producing high quality polymer components that are free from defects and residual stress, while accommodating increasingly complex geometries and large-scale fabrication. In this context, the development of predictive modeling frameworks that can accurately capture the underlying physics of polymerization becomes critical.

Polymerization is inherently a coupled process involving close interplays between thermodynamics, chemical kinetics and mechanical response. On the one hand, extensive experimental and modeling efforts have focused on thermo-chemical coupling, showing that reaction kinetics are strongly temperature-dependent \citep{yousefi1997kinetic,hardis2013cure}, and thus that the heat released during the exothermic polymerization reaction can influence the temperature distribution within the material. On the other hand, the polymerization process is accompanied by volumetric changes resulting from the competing effects of thermal expansion and chemical shrinkage. Spatially non-uniform heating, either from external sources or due to uneven reaction or cooling rates, could give rise to thermal gradients that generate residual stresses in components \citep{pusatcioglu1980effect, antonucci2002methodology}.
In applications such as mold-based fabrication \citep{guevara2014residual,yuan2016analytical}, protective coatings \citep{francis2002development}, and fiber-reinforced polymers \citep{zhao2006micromechanical,kablov2021influence}, volume mismatches between the polymer and surrounding materials (e.g., molds, substrates, or fibers) can induce significant residual stresses during curing. These stresses can compromise geometrical accuracy, degrade the mechanical performance and durability of the final component \citep{heinrich2013role}, and in severe cases, lead to warpage, interfacial delamination, or premature failure \citep{wang2023shrinkage}. Motivated by these challenges, recent efforts have aimed to develop predictive models that capture the thermo-chemo-mechanical coupling behavior in polymerization systems \citep{wu2018evolution,sain2018thermo,alkhoury2025chemo}.

Commonly used polymerization methods, such as thermal curing and photo-polymerization \citep{bagheri2019photopolymerization}, require continuous external energy input, such as heat, pressure, or light, throughout the entire process to sustain the polymerization reaction. As a result, these methods often rely on specialized equipment, such as autoclaves, which pose significant challenges for large-scale fabrication due to their high energy consumption and limited scalability \citep{abliz2013curing}. To overcome these limitations, frontal polymerization (FP) has recently emerged as a promising alternative, offering improved energy and time efficiency in the manufacturing of polymer components \citep{suslick2023frontal}. Instead of relying on  external energy  input,  FP utilizes the heat released from the exothermic polymerization reaction itself to sustain further reaction in adjacent regions. This self-sustained process produces a sharply localized front that separates the uncured and cured material and propagates autonomously, in contrast to bulk polymerization where the reaction progresses simultaneously throughout the domain. The propagation dynamics observed in FP are closely related to those in the combustion of solid fuels, whose theoretical foundation dates back to the classical studies of Zeldovich \citep{zeldowitsch1988theory}. Pulsating fronts have been reported in FP \citep{pojman1995spin, lloyd2021spontaneous} when uniform propagation loses stability. Following the theoretical developments in combustion \citep{matkowsky1978propagation}, the Zeldovich number has been adopted as a criterion for predicting the onset of such instabilities in FP \citep{solovyov1997numerical,lloyd2021spontaneous}.

 A key characteristic of FP is the presence of a sharp gradient zone at the front, where temperature and degree of curing change rapidly, which could lead to residual stress \citep{chen2024residual} or even cracks \citep{binici2006spherically, adrewie2025thiol} during the FP process, specifically for its large scale applications. At the same time, the rapid, localized, and self-sustaining nature of FP also makes it a compelling platform for investigating the coupled thermal, chemical, and mechanical behaviors that emerge during the polymerization process.

One fundamental coupling mechanism in polymerization systems that has not been included in existing frameworks is the influence of the mechanical stress state on the local reaction kinetics.  Polymerization  can be viewed as a  solid–solid phase transformation, similar to martensitic transformations \citep{abeyaratne1993continuum} and biological growth \citep{abi2019kinetics}. In these systems, the phase transformation behavior is known to be modulated by stress, as theoretically predicted and experimentally observed in contexts such as fracture extension \citep{rice1978thermodynamics}, and confined growth \citep{li2022nonlinear}. However, both theoretical and experimental investigations into such coupling effect during polymerization process remain limited. Recently, \citet{alkhoury2024investigating} demonstrated that applied stress can influence chain scission reactions during photo-degradation. In the context of frontal polymerization, our recent work \citep{li2024mechanical} provides the first direct experimental evidence that tensile stress can slow down, or even quench, the propagation of the polymerization front in a uniaxial stretch setup. Motivated by these observations, in this work we develop a thermodynamically consistent framework in which the local stress state enters the reaction kinetics through a driving force consistent with the second law of thermodynamics. Specifically for the case of FP, we further formulate a theoretical analysis of the propagation dynamics, including the propagation speed and the stability condition, and investigate how these features are influenced by mechanical loading.

Another important coupling mechanism contributing to the mechanical response during polymerization is the evolution of the material’s stress-free configuration as the reaction progresses. After the gel point, newly formed polymer chains become entangled and crosslinked with the existing network, leading to a increase in material stiffness. A key experimental observation across various polymerization systems \citep{gillen1988effect} is that if the material is deformed during the curing process, the deformation is not recovered, even after external loads are removed and polymerization is complete. This plastic-like behavior is commonly explained at the microscale by the assumption that newly formed polymer chains are deposited in a locally stress-free state. Several theoretical models have been developed to capture the evolution of this stress-free configuration. \cite{hossain2009small} adopted a hypoelastic formulation in which stress and strain are related incrementally in a rate form. \cite{wu2018evolution} employed a multibranch viscoelastic model to represent the stiffening during polymerization. More recently, \cite{kumar2022surface} proposed a phenomenological approach that approximates the behavior by assuming that the deformation at the instant of polymerization becomes permanently `frozen' into the material, and \cite{wijaya2025thermo} proposed an additive decomposition of the strain tensor and defined an inelastic strain to describe the evolution of the stress-free configuration.  In this work, inspired by flow plasticity theory, we adopt a multiplicative decomposition of the deformation gradient, which is more appropriate for large deformations and introduces a transformation deformation gradient with an associated evolution law, analogous to a flow rule, to describe the progression of the stress-free configuration. The formulation is developed in a thermodynamically consistent manner and is readily suited for finite element implementation.

The primary objective of this paper is to develop a large-deformation, thermo-chemo-mechanically coupled constitutive framework for modeling polymerization process, and to use it to elucidate the coupling mechanisms observed in frontal polymerization (FP). The manuscript is organized as follows: in Section 2, we  construct the fully coupled thermodynamically consistent theoretical framework for modeling the general polymerization process, with special focus on the stress influenced chemical kinetics and transformation deformation. Next, we apply the general framework to the case of FP under uniaxial loading, which enables analytical investigation. In Section 3, we focus on the propagation dynamics, including the propagation velocity and stability, and examine how mechanical loading and heat loss influences these behaviors. In Section 4, we analyze the force response under displacement control as the front propagates, which provides insight into the development of residual stress during the FP process. Finally, we offer some concluding remarks in Section 5.

\section{Theory}

\subsection{Kinematics}
\noindent
Consider a body $\mathcal{B}$ undergoing shape changes during a polymerization reaction. The body  occupies a fixed region $\Omega$ in its reference configuration, where $\bm{X}$ represents an arbitrary material point within $\mathcal{B}$. The motion of a material point in $\mathcal{B}$ at time $t$ is described by a smooth one-to-one mapping, $\bm{x} =\bm{\chi}(\bm{X},t)$, which relates the fixed region $\Omega$ in the reference configuration to the deformed region $\Omega_t$ in the current configuration. Accordingly, the deformation gradient $\bm{F}$, the velocity $\bm{v}$, and the velocity gradient $\bm{L}$ can be defined as\footnote{Note that here we use $\nabla$ and $\mathrm{Div}$ to denote the gradient and divergence with respect to the material point in the reference configuration, and use $\mathrm{grad}$ and $\mathrm{div}$ to denote the gradient and divergence in the current configuration. We use the superimposed dot to denote the material time derivative.}:

\begin{equation}
    \bm{F} = \nabla{\bm{\chi}},\quad \bm{v} = \dot{\bm\chi},\quad \bm{L} = \mathrm{grad}\ \bm{v} = \dot{\bm{F}} \bm{F}^{-1}.
\end{equation}

\begin{figure} [H]
    \centering
    \includegraphics[width=0.75\columnwidth]{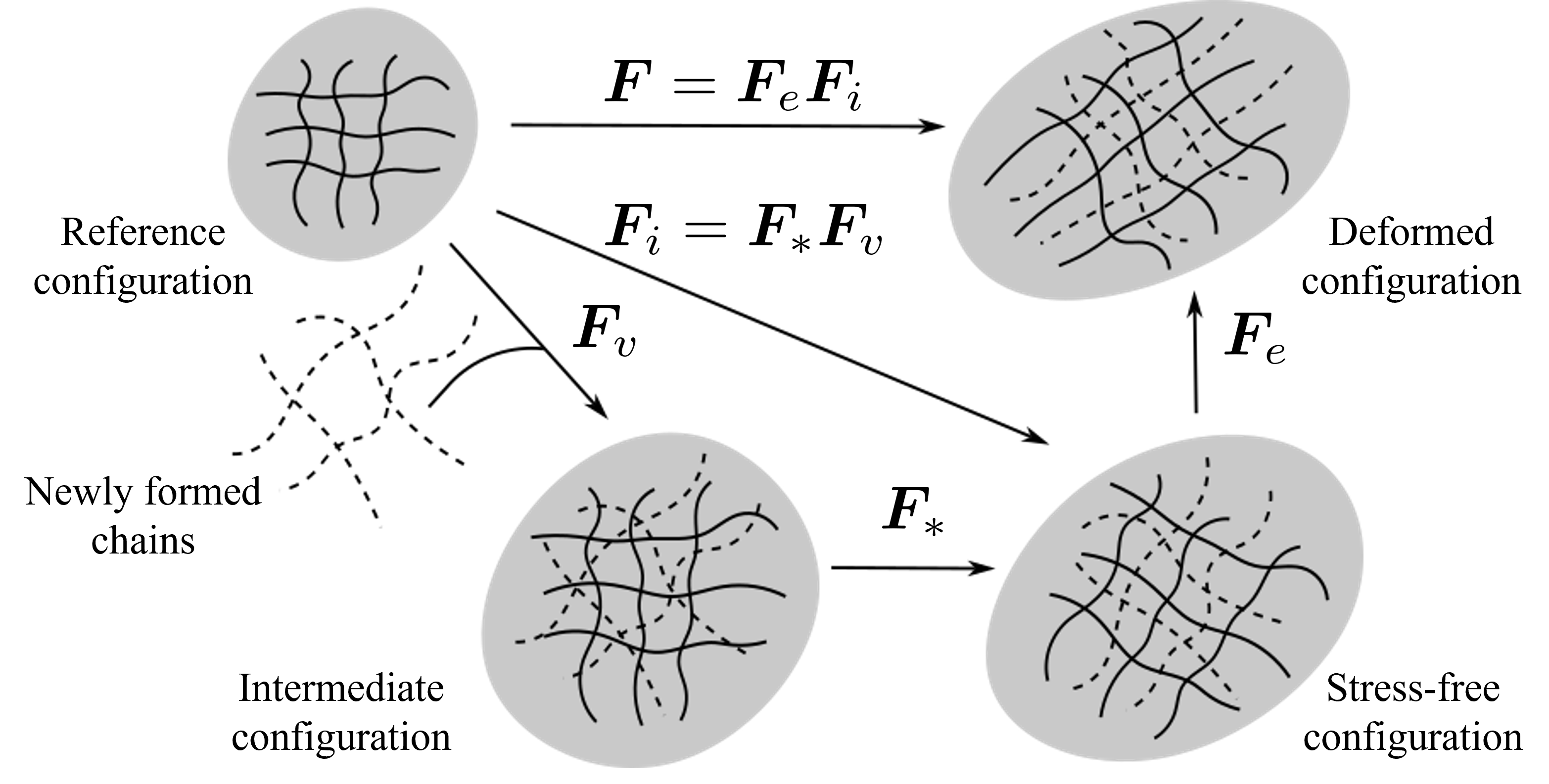}
    \caption{ Illustration of assumed kinematics via multiplicative decomposition of the deformation gradient $\bm{F} = \bm{F}_e\bm{F}_i$ where $\bm{F}_e$ is the elastic deformation gradient and $\bm{F}_i$ is the inelastic deformation gradient. The inelastic deformation gradient $F_i$ can be further decomposed as $\bm{F}_i = \bm{F}_*\bm{F}_v$, where $\bm{F}_v$ is a spherical tensor that represents the volume change, and $\bm{F}_*$ is the transformation deformation gradient.}
    \label{fig:kinematics}
\end{figure}

The deformation of the body ${\mathcal{B}}$ involves stretching and rotation of a polymer network relative to a local stress-free state, along with the evolution of this stress-free state relative to the reference configuration. Consequently, the deformation gradient can be multiplicatively decomposed into an elastic part and an inelastic part, as illustrated in Fig.\ref{fig:kinematics}:
\begin{equation}
\label{decomp_total}
    \bm{F} =\bm{F}_e\bm{F}_i,
\end{equation}
where $\bm{F}_e$ is the elastic distortion that represents the stretch and rotation of the elastic polymer network, and $\bm{F}_i$ is the inelastic deformation gradient that represents the evolution of the local stress-free state of the polymer network.

The decomposition of the deformation gradient in \eqref{decomp_total} introduces an intermediate state associated with the evolution of the material structure (i.e. the polymer network). Following the framework of finite plasticity theory \citep{gurtin2010mechanics}, we refer to  this intermediate state as the structural configuration, as it reflects the structure of the newly formed polymer network. 

In the context of the polymerization process, the evolution of the structural configuration can be attributed to two mechanisms: (a) volume changes due to thermal expansion and chemical shrinkage, and (b) isochoric distortion caused by  the deformation of the newly-formed polymer network under mechanical loading. Accordingly, the inelastic deformation gradient $\bm{F}_i$ is further decomposed as shown in Fig. \ref{fig:kinematics}:
\begin{equation}
\label{decomp_inelastic}
    \bm{F}_i = \bm{F}_*\bm{F}_v,
\end{equation}
where $\bm{F}_v = J_i^{1/3} \bm{1}$ is a spherical tensor with $J_i$ representing the volume change ratio, and $\bm{F}_*$ is the transformation deformation gradient, which is assumed to be isochoric, as is the elastic deformation gradient $\bm{F}_e$. As a result,  we have the following kinematic constraints on $\bm{F}_e$ and $\bm{F}_*$:
\begin{equation}
\label{incompressible_condition}
    \mathrm{det}(\bm{F}_e) = 1, \quad \mathrm{det}(\bm{F}_*) =1.
\end{equation}
The volumetric ratio of the deformation can then be written as:
\begin{equation}
    J_i = \mathrm{det}(\bm{F}) =  \mathrm{det}(\bm{F}_i).
\end{equation}
This relation reflects our kinematic assumption that the volume change arises solely from thermal expansion and chemical shrinkage.

Based on the decomposition relations \eqref{decomp_total} and \eqref{decomp_inelastic}, the velocity gradient $\bm{L} =\dot{\bm{F}}\bm{F}^{-1}$ can be decomposed as follows:
\begin{equation}
\label{velocity_decomp}
    \bm{L} = \bm{L}_e +\bm{F}_e\bm{L}_*\bm{F}_e^{-1} +\frac{1}{3} J_i^{-1} \dot{J}_i \bm{1},
\end{equation}
where $\bm{L}_e =\dot{\bm{F}}_e\bm{F}_e^{-1}$ is the elastic velocity gradient and $\bm{L}_* =\dot{\bm{F}}_*\bm{F}_*^{-1}$ is the transformation velocity gradient.

The velocity gradient $\bm{L}$ can be additively decomposed into the symmetric part $\bm{D}$ (stretching) and the skew-symmetric part $\bm{W}$ (spin):
\begin{equation}
\bm{L} = \bm{D} + \bm{W}, \quad
\bm{D} = \frac{1}{2}(\bm{L} + \bm{L}^\mathrm{T}), \quad
\bm{W} = \frac{1}{2}(\bm{L} - \bm{L}^\mathrm{T}).
\end{equation}

Similarly, the elasticity velocity gradient $\bm{L}_e$ and the transformation velocity gradient $\bm{L}_*$ can also be decomposed as:
\begin{equation}\label{symm}
    \begin{aligned}
        \bm{D}_e=\frac{1}{2}(\bm{L}_e+\bm{L}_e^T),\quad 
        \bm{W}_e=\frac{1}{2}(\bm{L}_e-\bm{L}_e^T),\\
                \bm{D}_*=\frac{1}{2}(\bm{L}_*+\bm{L}_*^T),\quad 
        \bm{W}_*=\frac{1}{2}(\bm{L}_*-\bm{L}_*^T),\\
    \end{aligned}
\end{equation}
so that we have $\bm{L}_e = \bm{D}_e +\bm{W}_e$, and $\bm{L}_* = \bm{D}_*+\bm{W}_*$.

The incompressibility condition \eqref{incompressible_condition} imposes the following constraint on the elastic strain rate and  transformation strain rate tensors:
\begin{equation}
    \mathrm{tr}(\bm{D}_e) =0, \quad\mathrm{tr}(\bm{D}_*) = 0.
\end{equation}
For the transformation spin tensor $\bm{W}_*$, a commonly used assumption in isotropic finite plasticity theory is that the plastic flow is irrotational\footnote{A detailed discussion on this assumption can be found in \cite{gurtin2005decomposition}.} (the plastic spin tensor vanishes). Following this approach, we adopt a similar kinematic assumption here: the transformation spin tensor $\bm{W}_*$ should vanish
\begin{equation}
\label{irrotational}
    \bm{W}_* =\bm{0}.
\end{equation}

\subsection{Balance Laws}
\subsubsection*{a) Mass conservation}
\noindent  We define $\rho(\bm{x},t)$ as the density field of the deformed body. In this work we neglect mass convection, hence  mass conservation requires:
\begin{equation}
    \frac{\dd}{\dd t} \int_{\Omega_{t}} \rho \dd V = 0,
\end{equation}
which in its local form implies:
\begin{equation}
\label{mass_local}
    \dot\rho + \rho \frac{\dot{J_i}}{J_i} =0.
\end{equation}

By integration of the above formula and setting $\rho_0$ as the initial density, we obtain the relation:
\begin{equation}
    \rho = \frac{\rho_0}{J_i}.
\end{equation}

\subsubsection*{b) Mechanical Equilibrium}
\noindent Given that the time scale associated with the polymerization process is significantly longer than that of inertial effects, we assume that the system remains in a state of mechanical equilibrium, which can be expressed in terms of the Cauchy stress tensor $\bm{T}$ as:

\begin{equation}
\label{mechanical equilibrium}
    \mathrm{div}({\bm{T}}) +\bm{b}  =\bm{0}, \quad \bm{T}^T =\bm{T},
\end{equation}
where $\bm{b}$ is a body force. The first equation represents the balance of linear momentum, and the second equation represents the balance of angular momentum.

\subsection{Energy Balance}
\noindent The polymerization process involves the conversion of energy among various forms, including internal energy, mechanical work, and heat. For the occupied region $\Omega_t$ in the deformed body with boundary $\partial\Omega_t$, the balance of energy can be written as\footnote{Here, $\dd V$ and $\dd A$ denote volume and surface area elements in the deformed configuration, respectively.}:
\begin{equation}\label{energy_balance_int}
\begin{aligned}
    \frac{\dd}{\dd t}{\int_{\Omega_t} {\rho\varepsilon} \dd V} &=\int_{\partial\Omega_t} \bm{T}\bm{n}\cdot\bm{v} \dd A  -\int_{\partial \Omega_t} \bm{q}\cdot\bm{n} \dd A + \int_{\Omega_t} \rho r \dd V,\\
\end{aligned}
\end{equation}
where $\varepsilon$ is the internal energy density per unit mass, $\bm{n}$ is the outward-facing unit normal to the surface, $\bm{t}=\bm{T}\bm{n}$ therefore represents the surface traction, $\bm{q}$ is the heat flux, and $r$ is the external heat supply per unit mass\footnote{In most cases, the dominant contribution to $r$ arises from heat loss to the environment. The form of this heat loss depends on the dissipation mechanism. For radiative cooling, the heat loss may be modeled using the Stefan–Boltzmann law, yielding $r \propto (T^{4} - T_{0}^{4})$. For convective cooling, Newton’s law of cooling gives $r \propto (T - T_{0})$. Alternatively, heat dissipation may be imposed directly through a boundary condition on the temperature field rather than introduced as a source term.}. Note that in the present framework the internal energy $\varepsilon$ is taken to include the chemical energy associated with the polymerization reaction. As a result, the heat released by the reaction is accounted for through the change in internal energy and does not appear explicitly as an additional source term in the energy balance.

Since \eqref{energy_balance_int} applies to any subregion of $\Omega_t$, by applying the divergence theorem and using the force balance \eqref{mechanical equilibrium},  the localized energy balance equation reads:
\begin{equation}
\label{energy_balance}
    \rho\dot{\varepsilon} -\bm{T}:\bm{L} + \mathrm{div}\ \bm{q} -\rho r =0.
\end{equation}
\subsection{Entropy imbalance}
Considering an irreversible process in an open system, the second law of thermodynamics implies that the total entropy never decreases, namely:
\begin{equation}
    \frac{d}{dt}\int_{\Omega_t} \rho{S} \dd V \geq -\int_{\partial \Omega_t} \frac{\bm{q}\cdot\bm{n}}{T} \dd A +\int_{\Omega_t}\frac{\rho r}{T} \dd V,
\end{equation}
where $S$ is the entropy per unit mass, and $T$ is the temperature field. This can be localized to give the Clausius–Duhem inequality:
\begin{equation}
\label{Clausius}
    \rho\dot{S}\geq -\mathrm{div}\left(\frac{\bm{q}}{T}\right) +\frac{\rho r}{T}.
\end{equation}
For the coupled system, we introduce the Helmholtz free energy per unit mass $\psi$:
\begin{equation}\label{Helmholtz}
    \psi = \varepsilon - T\cdot S,
\end{equation}
we can then use \eqref{energy_balance} and \eqref{Helmholtz} to rewrite \eqref{Clausius}  and write the free energy imbalance:
\begin{equation}
\label{imbalance}
    \rho\dot{\psi} - \bm{T}:\bm{L} +\rho S\cdot\dot{T}  +\frac{1}{T}\bm{q}\cdot\bm{\nabla}T\leq 0.
\end{equation}

Note that the stress power $\bm{T}:\bm{L}$ appears in both the energy balance equation \eqref{energy_balance} and the free energy imbalance \eqref{imbalance}, representing the mechanical power. Utilizing the decomposition of the velocity gradient \eqref{velocity_decomp} and the irrotational condition \eqref{irrotational}, we can decompose the stress power as follows:
\begin{equation}\label{power_decomp}
\begin{aligned}
    \bm{T}:\bm{L} &= \bm{T}:\left(\bm{L}_e+ \bm{F}_e\bm{L}_*\bm{F}_e^{-1} +\frac{1}{3}J_i^{-1} \dot{J}_i\bm{1}\right)\\
    & = \frac{1}{2} (\bm{F}_e^{-1}\bm{T}\bm{F}_e^{-T}):\dot{\bm{C}}_e + (\bm{F}_e^T\bm{T}\bm{F}_e^{-T}):\bm{D}_* - J_i^{-1}p\cdot \dot{J}_i,
\end{aligned}    
    \end{equation}
where $\bm{C}_e = \bm{F}_e^T\bm{F}_e$ is the right Cauchy-Green strain tensor, and $p = -\ \mathrm{tr}(\bm{T})/3$  is the hydrostatic pressure.

Accordingly, the mechanical power can be partitioned into three distinct contributions, following the kinematic decomposition of the total deformation. Specifically, in \eqref{power_decomp} the first term corresponds to the mechanical power associated with the elastic deformation, the second term corresponds to the transformation deformation, and the last term is related to the volume change. This decomposition of stress power motivates the introduction of two additional stress measurements based on the Cauchy stress $\bm{T}$. The second Piola stress $\bm{S}$ defined as:
\begin{equation}
\label{second_piola}
    \bm{S} = \bm{F}_e^{-1}\bm{T}\bm{F}_e^{-T},
\end{equation}
which is conjugate to $\dot{\bm{C}}_e$, and the Mandel stress $\bm{M}$:
\begin{equation}\label{mandel}
    \bm{M} = \bm{F}_e^T\bm{T}\bm{F}_e^{-T},
\end{equation}
which is conjugate to $\bm{D}_*$.

\subsection{Constitutive relations}
\noindent To develop the thermodynamically consistent constitutive relations that satisfy the free energy imbalance \eqref{imbalance}, we  follow the Coleman-Noll methodology \citep{noll1974thermodynamics}. 
We assume that the energetic state of the polymerization system is characterized by its elastic deformation (expressed through $\bm{C}_e$), its temperature $T$, and the degree of curing, which we denote by $\alpha$. Following the approach in \cite{li2024mechanical}, we also feature another energetic contribution arising from the stored energy within the polymer network during the irreversible reaction. Accordingly, the Helmholtz free energy is expressed as:
\begin{equation}
\label{decomp_defect}
    \psi = {\psi}_r(\bm{C}_e,T,\alpha) +\psi_*,
\end{equation}
where $\psi_r$ is the reversible part of the free energy, and $\psi_*$  accounts for the irrecoverable energy stored in the polymer network and thus depends on the history (similar to the concept of the defect energy 
in plasticity theory).

Based on the form of the free energy in \eqref{decomp_defect}, we can write the rate of change of the free energy as:
\begin{equation}
\label{chain_rule}
    \dot\psi =  \frac{\partial \psi_r}{\partial \bm{C}_e} :\dot{\bm{C}}_e +\frac{\partial \psi_r}{\partial T}\dot{T} + \frac{\partial \psi_r}{\partial \alpha}\dot{\alpha} +\dot{\psi}_*,
\end{equation}

Substituting  \eqref{power_decomp}  and \eqref{chain_rule} into the free energy imbalance \eqref{imbalance}, the thermodynamic process satisfies the entropy condition if and only if the following inequality holds:
\begin{equation}
\label{major}
    \left(\rho\frac{\partial \psi_r}{\partial \bm{C}_e}-\frac{1}{2}\bm{S}\right):\dot{\bm{C}}_e +\left(\rho\frac{\partial \psi_r}{\partial T}+\frac{p}{J_i} \frac{\partial J_i}{\partial T}+\rho S\right)\dot{T} +\frac{1}{T}\bm{q}\cdot\bm{\nabla}T + \left(\rho\frac{\partial \psi_r}{\partial \alpha} + \frac{p}{J_i}\frac{\partial J_i}{\partial \alpha}\right) \dot\alpha + (\rho\dot{\psi}_i-\bm{M}:\bm{D}_*) \leq0,
\end{equation}
which should hold for any arbitrary combination of the rates $\dot{\bm{C}}_e, \dot{T},\dot{\alpha}$, $\bm{\nabla} T$, $\bm{D}_*$. As a result, each term in \eqref{major} should either vanish or be negative, as it characterizes a distinct dissipative mechanism in the chemo-thermo-mechanically coupled system: the first and  second terms represent  rate-dependent behaviors of the mechanical and thermal response of the material; the third term describes entropy generation during heat conduction; the fourth term is associated with the dissipation due to polymerization reaction, and the last term is the heat generated by the transformation deformation.

In the present study, we neglect any rate-dependent effects in the material response, such as viscoelasticity. Consequently, the first and second terms must vanish. Considering the incompressibility condition described in \eqref{incompressible_condition} (with a detailed derivation provided in \ref{app:incomp}), the constitutive relation for the Cauchy stress, based on the first term in \eqref{major}, is:
\begin{equation}
\label{cons_stress}
    \bm{T} = 2\rho\ \mathrm{dev}\left(\bm{F}_e\frac{\partial \psi_r}{\partial \bm{C}_e}\bm{F}_e^T\right) -p\bm{1},
\end{equation}
where the pressure field $p$ is constitutively indeterminate and must be obtained from the specified traction boundary conditions.

Similarly, the second term in \eqref{major} should vanish, which provides the constitutive relation for the entropy $S$:
\begin{equation}
\label{cons_entropy}
    S = -\frac{\partial \psi_r}{\partial T} - \frac{p}{\rho_0}\frac{\partial J_i}{\partial T}.
\end{equation}
The last three terms in \eqref{major} contribute to dissipative processes. Specifically, the third term implies that:
\begin{equation}\label{heat_ineq}
    \bm{q} \cdot\bm{\nabla}T\leq 0,
\end{equation}
and the fourth term can be rewritten by  introducing the driving force for the curing process:
\begin{equation}
\label{cons_driving}
    \mathcal{F}(\bm{C}_e,T,\alpha) = -  \frac{\partial\psi_r}{\partial\alpha} - \frac{p}{\rho_0} \frac{\partial J_i}{\partial \alpha},
\end{equation}
The free energy imbalance implies the inequality:
\begin{equation}\label{cons_chem}
    \mathcal{F}\cdot\mathcal{G} \ge 0,
\end{equation}
where $\dot{\alpha} = \mathcal{G}(\bm{C_e},T,\alpha)$.

The last term in \eqref{major} imposes the requirement that the increase in the network energy $\psi_*$ must always be less than or equal to the mechanical work invested in the transformation deformation:

\begin{equation}\label{network_energy}
\dot{\psi}_* \leq \bm{M}:\bm{D}_*, 
\end{equation}
where the difference between these two terms corresponds to the heat generated due to transformation deformation. Furthermore, considering the irreversible nature of the network energy $\psi_*$, we impose the condition $\dot{\psi}_* \geq 0$, which leads to the additional constraint that the mechanical work invested in transformation deformation must always be non-negative:

\begin{equation}\label{cons_flow} 
\bm{M}:\bm{D}_* \geq 0.
\end{equation}
This constraint imposes a restriction on the evolution relation of the transformation deformation $\bm{D}_*$, which is similar to the flow rule.

In summary, based on the second law of thermodynamics \eqref{major}, we establish the constitutive relations for the Cauchy stress in \eqref{cons_stress} and the entropy in \eqref{cons_entropy}, and have obtained restrictions on the constitutive response functions for the heat flux in \eqref{heat_ineq}, the chemical kinetics in \eqref{cons_chem}, and the flow rule in \eqref{cons_flow}. 

\subsection{Constitutive response functions}
So far, our discussion has remained general and applicable to a broad class of systems involving irreversible chemical reactions. In this section, we focus on defining the constitutive relations specific to the polymerization system. To achieve this, we first choose the explicit form of the free energy function $\psi_r$ in a simple separable manner:

\begin{equation} \label{response} \psi_r(\bm{C}_e,T,\alpha) = \frac{\mu(\alpha)}{2\rho}\left(\mathrm{tr}(\bm{C}_e)-3\right) +\tilde{H}(1-\alpha) + \tilde{c}\left[\left(T-T_0\right)-T \mathrm{ln}\left(\frac{T}{T_0}\right)\right], \end{equation}
where the first term represents the neo-Hookean elastic energy density per unit mass, 
the function $\mu(\alpha)$ denotes the curing-dependent shear modulus, which can increase significantly as the polymerization process progresses.
To capture the chemical hardening effect that occurs during curing, we adopt a simple linear relation for $\mu(\alpha)$:
\begin{equation} \label{chemical_hardening}
    \mu(\alpha) = \mu_0 +\mu_H\alpha,
\end{equation}
where $\mu_0$ represents the initial shear modulus in the gel state (i.e., $\alpha = 0$), and $\mu_H$ is the chemical hardening modulus \footnote{For frontal polymerization of dicyclopentadiene (DCPD) starting at a gelled state, the stiffness contrast is significantly large, with $\mu(\alpha=1)/\mu(\alpha=0)\sim 10^5$.}.

The second term in the free energy expression \eqref{response} represents the chemical energy, where $\tilde{H}$ denotes the chemical modulus per unit mass. The third term accounts for the thermal energy, with $\tilde{c}$ representing the heat capacity per unit mass. 

For the kinetic description of the volume change ratio as a function of temperature $T$ and the degree of curing $\alpha$, we adopt the following linear relation:
\begin{equation} \label{volume}
J_i(T,\alpha) = 1 + \beta T - \gamma \alpha, \end{equation}
where $\beta$ is the thermal expansion coefficient, and $\gamma$ represents the chemical shrinkage coefficient.

Substituting \eqref{response} into \eqref{cons_stress}, \eqref{cons_entropy} and \eqref{cons_driving}, we write the relations for the Cauchy stress $\bm{T}$, the entropy $S$ and the driving force $\mathcal{F}$:
\begin{equation} \label{const}
\begin{aligned}
    \bm{T}& = \mu(\alpha)\cdot\mathrm{dev}(\bm{B}_e)-p\bm{1},\\
    S &= \tilde{c}\cdot \ln\left(\frac{T}{T_0}\right) - \frac{\beta}{\rho_0} p,\\
    \mathcal{F} &= \tilde{H} -\frac{1}{2\rho}\frac{d\mu}{d\alpha}(\mathrm{tr}(\bm{C}_e)-3) +   \frac{\gamma}{\rho_0}p,
\end{aligned}
\end{equation}
where $\bm{B}_e = \bm{F}_e \bm{F}_e^T$ is the left Cauchy-Green tensor.

Next, for the heat flux, to ensure that the condition in \eqref{heat_ineq} is always satisfied, we adopt the commonly used Fourier-type law:
\begin{equation}
    \bm{q} = - \kappa \bm{\nabla} T,
\end{equation}
where $\kappa >0$ is the thermal conductivity.

To maintain thermodynamic consistency in the chemical kinetics and ensure that the constraint in \eqref{cons_chem} is always satisfied, we follow \cite{li2024mechanical} and introduce the reaction rate as:
\begin{equation}\label{chem_kinetic}
    \mathcal{G}(\bm{C}_e,T,\alpha) = \left\{
\begin{aligned}
     Ce^{-\frac{E_a}{RT}}(1-\alpha)^n\left(\frac{\mathcal{F}(\bm{C}_e)-\mathcal{F}_0(T)}{\tilde{H}}\right)&\ &\mathcal{F}\ge \mathcal{F}_0 \\
    0\qquad\qquad&\ &\mathcal{F}< \mathcal{F}_0,\\
\end{aligned}    
\right.
\end{equation}
where $C$ is the reaction rate constant, $E_a$ is the chemical activation energy, $n$ is the chemical reaction order and $\mathcal{F}_0$ is the energy barrier given by:
\begin{equation}\label{barrier}
\mathcal{F}_0(T) =   \Gamma_0 \frac{E_a}{RT}
\end{equation}
which accounts for the influence of the local temperature $T$
with  magnitude coefficient $\Gamma_0$. 

\subsection{Evolution of transformation deformation}
\noindent To complete the theoretical framework, it is necessary to establish a kinetic relation governing the evolution of transformation deformation that is consistent with the thermodynamic constraint \eqref{cons_flow}. The evolution of transformation deformation can be fully determined by a kinetic description of $\bm{D}_*$. An analogy can be drawn between this evolution relation and the flow rule in plasticity theory, which governs the evolution of plastic deformation. In plasticity, the flow rule is derived from a consistency condition that ensures the stress state remains on the yield surface as plastic flow occurs. Similarly, in the context of the polymerization process, the evolution of $\bm{D}_*$ should be governed by a consistency condition that reflects the fundamental physics of the polymerization process while ensuring that the corresponding material response remains thermodynamically and mechanically admissible.

The curing process directly influences the mechanical response of the material in two ways: (i) volume change due to thermal expansion and chemical shrinkage, and (ii) an increase in material stiffness resulting from the formation of new polymer chains within the existing network. While the first effect has already been incorporated into the framework through a direct kinematic description of the volume change ratio $J_i$ in \eqref{volume}, the influence of the chemical hardening effect on mechanical behavior requires additional considerations. To achieve this, we need to examine the mechanical response during polymerization in rate form.

Let $t_0$ be an arbitrarily chosen but fixed time at which the fields of motion $\bm{\chi}$, the elastic deformation gradient $\bm{F}_e$, and the transformation deformation gradient $\bm{F}_*$, are known. At $t=t_0$, we introduce the corresponding virtual fields: the velocity $\tilde{\bm{v}}$, the elastic velocity gradient $\tilde{\bm{L}}_e$, and the transformation velocity gradient $\tilde{\bm{L}}_*$, which must obey the kinematic relation \eqref{velocity_decomp}:
\begin{equation}\label{virtural_velocity}
    \mathrm{grad}\ \tilde{\bm{v}} =\tilde{\bm{L}}= \tilde{\bm{L}}_e +\bm{F}_e\tilde{\bm{L}}_*\bm{F}_e^{-1},
\end{equation}
The tilde notation is used to distinguish these fields from those corresponding to the actual motion. These virtual fields do not represent the true evolution of the body but rather kinematically admissible variations of the motion. The objective is to use these virtual velocity fields to derive a consistency condition that characterizes mechanically admissible motions during the curing process.

With the virtual velocity fields $\{\tilde{\bm{L}},\tilde{\bm{L}}_e,\tilde{\bm{L}}_*\}$, we can now write the constitutive relation of the deviatoric part of Cauchy stress ($\bm{T}_0 \equiv \mathrm{dev}(\bm{T})$)  \eqref{const}$^1$ in rate form:
\begin{equation}\label{rate_general} 
\dot{\bm{T}}_0 = \dot{\mu} \cdot \mathrm{dev}(\bm{B}_e) + \mu \cdot \mathrm{dev}(\tilde{\bm{L}}_e \bm{B}_e + \bm{B_e}\tilde{\bm{L}}_e^T), 
\end{equation}
where $\bm{B}_e = \bm{F}_e\bm{F}_e^T$. The two terms in equation~\eqref{rate_general} represent distinct contributions to the evolution of stress: the first term arises from chemical hardening due to the evolution of material stiffness, while the second term accounts for the change in elastic deformation.

We first investigate the case where no polymerization takes place. In this scenario, there is no chemical stiffening, i.e., $\dot{\mu} = 0$, and consequently, there is no evolution of the stress-free configuration ($\tilde{\bm{L}}_* = \bm{0}$). From the kinematic constraint \eqref{virtural_velocity}, this implies that the total virtual velocity gradient reduces to the elastic part, i.e., $\tilde{\bm{L}} = \tilde{\bm{L}}_e$. Substituting into \eqref{rate_general}, the stress evolution as a function of the virtual velocity gradient $\tilde{\bm{L}}$ in this pure-elastic case ($\tilde{\bm{L}}_* =\bm{0}$) is given by:
\begin{equation}\label{rate_special}
\dot{\bm{T}}_0 = \mu \cdot \mathrm{dev}(\tilde{\bm{L}} \bm{B}_e + \bm{B}_e \tilde{\bm{L}}^T).
\end{equation}

Now consider the general case for the evolution of stress during curing. As discussed previously, the formation of new polymer chains during the polymerization process leads to a significant increase in stiffness, i.e., $\dot{\mu} > 0$. To provide a physical explanation for why the stress-free configuration evolves during curing, a widely adopted assumption in the literature \citep{hossain2009small, wu2018evolution, wijaya2025thermo} is that newly formed polymer chains are deposited in a locally stress-free state at the moment of their formation. As a consequence, this change in stiffness should not directly contribute to the evolution of stress. 
To incorporate this assumption into our theoretical framework, we propose a consistency condition for the polymerization process: at each moment during curing, the stress should evolve in the same manner as it would in an elastic body with the same instantaneous stiffness but without any ongoing curing. This condition corresponds exactly to the stress evolution relation in pure-elastic case previously derived in \eqref{rate_special}.

The consistency condition implies that the expression for stress evolution derived for the case of zero transformation deformation rate in \eqref{rate_special} should also apply in the general case where $\bm{L}_* \neq \bm{0}$.  Combining the general constitutive relation \eqref{rate_general} with \eqref{rate_special}, we obtain:
\begin{equation}\label{eq:consistent}
   \dot{\mu} \cdot \mathrm{dev}(\bm{B}_e) + \mu \cdot \mathrm{dev}(\tilde{\bm{L}}_e \bm{B}_e + \bm{B_e}\tilde{\bm{L}}_e^T)  =\mu \cdot \mathrm{dev}(\tilde{\bm{L}} \bm{B}_e + \bm{B}_e \tilde{\bm{L}}^T).
\end{equation}
Substituting the kinematic constraint from \eqref{virtural_velocity} to eliminate $\tilde{\bm{L}}$ and $\tilde{\bm{L}}_e$, and applying the irrotational condition given in \eqref{irrotational}, we obtain the evolution equation for transformation strain rate tensor $\bm{D}_*$:
\begin{equation} \label{flow_v1} \mathrm{dev}\left( \bm{F}_e {\bm{D}_* \bm{F}_e^T} \right) = \frac{\mu_H \dot{\alpha}}{2\mu^2} \bm{T}_0,
\end{equation}
which implies that $\bm{D}_*$ has the following form:
\begin{equation}
    \bm{D}_* =   \frac{\mu_H \dot\alpha}{2\mu^2} \left(\bm{F}_e^{-1}\bm{T}_0\bm{F}_e^{-T} + \zeta \bm{F}_e^{-1}\bm{F}_e^{-T}\right),
\end{equation}
where $\zeta$ is a scalar to be determined. 

To determine the value of $\zeta$, we impose the incompressibility condition on the transformation deformation \eqref{incompressible_condition} that requires $\bm{D}_* = \bm{0}$, from which we can solve for $\zeta$:
\begin{equation}
    \zeta = - \frac{\mathrm{tr}(\bm{F}_e^{-1}\bm{T}_0\bm{F}_e^{-T})}{\mathrm{tr}(\bm{F}_e^{-1}\bm{F}_e^{-T})}.
\end{equation}

As a result, we have the evolution relation for the transformation deformation:
\begin{equation}\label{flow_rule}
       \bm{D}_* =   \frac{\mu_H \dot\alpha}{2\mu^2} \left(\bm{F}_e^{-1}\bm{T}_0\bm{F}_e^{-T} - \frac{\mathrm{tr}(\bm{F}_e^{-1}\bm{T}_0\bm{F}_e^{-T})}{\mathrm{tr}(\bm{F}_e^{-1}\bm{F}_e^{-T})}\bm{F}_e^{-1}\bm{F}_e^{-T}\right).
\end{equation}

The evolution relation \eqref{flow_rule} ensures that the material response predicted by our theoretical framework remains mechanically admissible. However, it is also necessary to verify whether the material response governed by \eqref{flow_rule} is thermodynamically admissible. Specifically, we must check whether the proposed evolution equation satisfies the thermodynamic inequality \eqref{cons_flow}. This verification has been carried out in \ref{app:flow_thermo}, where it is shown that \eqref{flow_rule} is indeed thermodynamically consistent.

It is important to note that the flow rule derived here is based on the specific form of the free energy function $\psi_r$ given in \eqref{response}. More generally, the derivation procedure outlined above can be extended to any given free energy function. A detailed derivation for the general case is provided in \ref{app:flow_general}.

Finally, we consider the energy conversion associated with the transformation deformation, as governed by the thermodynamic constraint \eqref{network_energy}, which requires that the heat generation due to the evolution of transformation deformation must be non-negative. Since there is no explicit energy dissipation mechanism associated with the evolution of transformation deformation, we further assume that there is no heat generation due to the evolution of transformation deformation, which indicates: 
\begin{equation} 
\dot{\psi}_* = \bm{M}:\bm{D}_*. 
\end{equation}

With this assumption, we can write the heat equation which governs the evolution of the temperature field based on the energy balance equation \eqref{energy_balance}. A detailed derivation is provided in \ref{app:heat}.
\begin{equation} \label{heat} \tilde{c} \dot{T} = \frac{1}{\rho}\bm{\nabla} \cdot (\kappa \bm{\nabla} T) + \mathcal{F} \dot{\alpha} + \frac{\beta T}{\rho_0}\dot{p} + r. \end{equation}

\subsection{Summary of governing equations}
As a summary, here we give the four governing equations which describe the evolution of the displacement $\bm{u}$, the temperature $T$, the degree of curing $\alpha$, and the transformation deformation $\bm{F}_*$:
\begin{itemize}

 \item[\textit{I}.] {Mechanical equilibrium} \eqref{mechanical equilibrium}:
\begin{equation}
\label{govern_1}
    \bm{\nabla}\cdot {\bm{T}} =\bm{0}.
\end{equation}

 \item[\textit{II}.] Heat balance \eqref{heat}:
\begin{equation}\label{govern_2}
    \tilde{c}\dot T = \frac{1}{\rho}\bm{\nabla}\cdot(\kappa\bm{\nabla}T)+ \mathcal F \dot\alpha +\frac{\beta T}{\rho_0}\dot{p} +r,
\end{equation}
with  
\begin{equation}\label{reaction_heat_full}
    \mathcal{F} =\tilde{H}  -\frac{\mu_H
    }{2\rho}(\mathrm{tr}(\bm{C}_e)-3) +   \frac{\gamma}{\rho_0}p,
\end{equation}
obtained from \eqref{const}.

 \item[\textit{III}.] Chemical kinetics:
\begin{equation}\label{govern_3}
    \dot\alpha = \hat{\mathcal{G}} (\bm{C}_e,T,\alpha),
\end{equation}
where the form of $\hat{\mathcal{G}}$ is written in \eqref{chem_kinetic}.

 \item[\textit{IV}.] Flow rule \eqref{flow_rule}
\begin{equation}\label{govern_4}
    \bm{D}_* = \frac{\mu_H}{2\mu^2}\left(\bm{F}_e^{-1}\bm{T}_0\bm{F}_e^{-T} - \frac{\mathrm{tr}(\bm{F}_e^{-1}\bm{T}_0\bm{F}_e^{-T})}{\mathrm{tr}(\bm{F}_e^{-1}\bm{F}_e^{-T})}\bm{F}_e^{-1}\bm{F}_e^{-T}\right)\dot{\alpha}.
\end{equation}
\end{itemize}

The theoretical framework established here captures the complex coupled behaviors inherent in general polymerization processes. Two distinct coupling mechanisms are featured: (i) the influence of local stress states on the reaction kinetics, explicitly incorporated through the chemical kinetic relation in \eqref{govern_3}, and (ii) the evolution of the stress-free configuration governed by the transformation deformation described by the flow rule \eqref{govern_4}. 

The formulation developed here is fully three-dimensional and applicable to problems with arbitrary geometries and boundary conditions. However, quantitative investigation of such systems typically requires numerical implementation which can obscure the underlying physics of the coupling. To gain physical insight and isolate the essential features of the coupled response, we next specialize the theory to one-dimensional configurations, which enables analytical investigation of the two coupling mechanisms.


\section{Coupling mechanism I: Propagation dynamics}
\label{section:dynamics}
\noindent
In this section, we investigate how the stress state influences the reaction kinetics under uniaxial stretch. We adopt a one-dimensional simplification, considering variations only along the primary propagation axis (denoted by coordinate $x$). To isolate the effect of stress on polymerization, we consider the process under a prescribed uniaxial stress, which is spatially uniform by equilibrium. This mechanical loading is equivalently characterized by the elastic stretch
$\lambda_e$ in the uncured region through the constutive relation \eqref{const}$^1$, which will be used throughout this section. A detailed discussion of the evolution of the mechanical field during polymerization process will be presented in Section 4; here, we neglect any evolution of the mechanical field during the process and assume $\lambda_e$ to be constant.

As a result, the problem reduces to the evolution of two coupled fields: the temperature $T(x,t)$ and the degree of curing $\alpha(x,t)$. Considering the heat balance equation~\eqref{govern_2} and the chemical kinetics relation~\eqref{govern_3}, and applying the above assumptions, we obtain the reduced reaction–diffusion system:
\begin{equation}\label{eq:reac_diffusion}
\left\{
\begin{aligned}
    &c\dot{T} = {\kappa} T_{xx} + \bar{\mathcal{F}}
    (\lambda_e) \cdot \mathcal{G}(T,\alpha,\lambda_e)-h(T-T_0),\\
    &\dot{\alpha} = \mathcal{G}(T,\alpha,\lambda_e).
\end{aligned}
\right.
\end{equation}
Here, it is convenient to rewrite the heat balance equation \eqref{govern_2} in terms of quantities per unit volume. To this end, we introduce the volumetric heat capacity $c =\rho\tilde c$ and the driving force per unit volume $\bar{\mathcal{F}} = \rho\mathcal{F}$. Note that the term associated with $\dot{p}$ in \eqref{govern_2} vanishes, since the pressure remains constant under the prescribed uniaxial stress. Heat loss to the environment is modeled using Newton’s cooling law $r = h/\rho\cdot(T-T_0)$, where $h$ denotes the dissipation coefficient and $T_0$ is the room temperature. 

The general form of the driving force $\mathcal{F}$ given in \eqref{reaction_heat_full} can be further simplified under the uniaxial loading condition as a function of the elastic stretch $\lambda_e$:
\begin{equation}\label{eq:driving_1d}
    \bar{\mathcal{F}}(\lambda_e) = {H}  - \frac{1}{2}\mu_H(\lambda^2_e+2\lambda_e^{-1} -3),
\end{equation}
where $H =\rho\tilde{H}$ is the reaction enthalpy per unit volume. The term involving the pressure $p$ in \eqref{reaction_heat_full} has been neglected, as it is much smaller in magnitude compared to the contribution arising from chemical hardening in the case of uniaxial stretch. 

Based on the expression of the driving force,  the thermodynamically consistent chemical kinetic relation $\mathcal{G}$ given in \eqref{chem_kinetic} is further specialized to:
\begin{equation}\label{chem_kinetic_1D}
    \mathcal{G}(T,\alpha,\lambda_e) = \left\{
\begin{aligned}
     &Ce^{-\frac{E_a}{RT}}\left(\frac{T_0}{T_*(\lambda_e)} - \frac{T_0}{T}\right)\cdot (1-\alpha)\ & T \geq T_*(\lambda_e), \\
    &0\qquad\qquad\ & T< T_*(\lambda_e),\\
\end{aligned}    
\right.
\end{equation}
where we have chosen the magnitude coefficient $\Gamma_0 = HRT_0/E_a$ for simplicity and take the order of chemical reaction $n =1$. In the general formulation~\eqref{chem_kinetic}, the reaction rate vanishes when the thermodynamic driving force $\mathcal{F}$ falls below the critical value $\mathcal{F}_0$. In the present setting, this criterion can be equivalently represented by introducing a cutoff temperature $T_*(\lambda_e)$, below which the driving force remains lower than the critical threshold and polymerization ceases:
\begin{equation}\label{eq:cut_off}
    T_*(\lambda_e) =T_0 \left(1-\frac{\mu_H}{2H}(\lambda_e^2+2\lambda_e^{-1}-3)\right)^{-1},
\end{equation}
noting that $T_*(\lambda_e)$ depends explicitly on the elastic stretch and  $T_*(\lambda_e=1) = T_0$.

At this stage, we have established a complete set of governing equations suitable for analyzing the propagation dynamics under prescribed mechanical conditions.
It turns out that the same formulation can lead to two distinct regimes of polymerization, depending on the material properties and boundary conditions: bulk polymerization and frontal polymerization.

\subsection{Characteristic regimes of polymerization}
\label{sec:regimes}
\noindent
Based on the governing equations~\eqref{eq:reac_diffusion} established in the previous section, we now examine the two characteristic regimes of polymerization.
In the first regime, known as bulk polymerization, the reaction is sustained by continuous external heating and progresses gradually through the material due to thermal diffusion.
In the second regime, known as frontal polymerization, a localized, self-sustained reaction front forms and propagates through the material once initiated.

\begin{figure} [h]
    \centering
    \includegraphics[width=1.0\columnwidth]{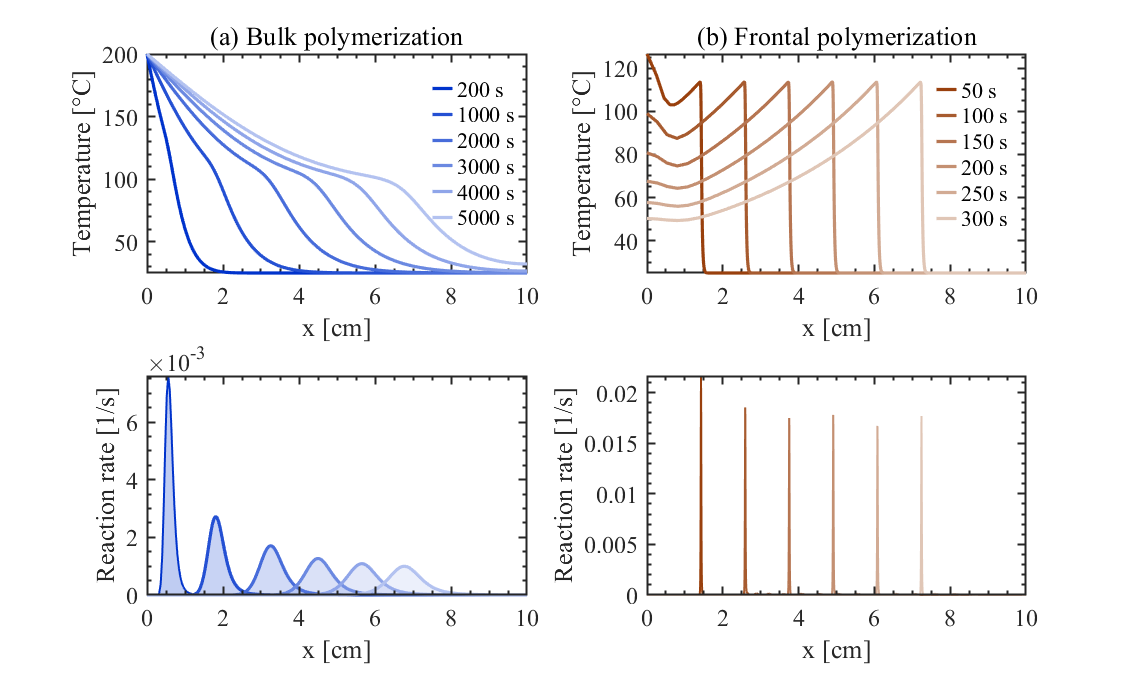}
\caption{Comparison between bulk and frontal polymerization regimes obtained from numerical simulations of the reduced one-dimensional governing equations~\eqref{eq:reac_diffusion}.
The top panels show the temporal evolution of the temperature profiles, while the bottom panels present the corresponding reaction-rate distributions $\dot{\alpha}(x,t)$, indicating the reaction zones where polymerization occurs.
(a) {Bulk polymerization,} where continuous heating is applied at $x = 0$;
(b) {Frontal polymerization}.}
    \label{fig:mode}
\end{figure}

To illustrate these two regimes, we numerically solve~\eqref{eq:reac_diffusion} using representative material parameters corresponding to each case.
For bulk polymerization, we consider a polydimethylsiloxane (PDMS)-based system, with parameters selected from~\cite{sain2018thermo}.
For frontal polymerization, we examine a dicyclopentadiene (DCPD)-based system, using parameters taken from~\cite{li2024mechanical}. The values of the material parameters used in the simulations are listed in \ref{app:values}. The resulting polymerization processes are shown in Fig.~\ref{fig:mode}.

For bulk polymerization, continuous external heating is required to sustain the reaction.
To simulate this condition, the temperature at the left boundary is fixed at $T(x=0)=200^{\circ}\mathrm{C}$, representing a constant heat source.
The temperature profiles show that heat gradually diffuses into the neighboring region, resulting in a slow temperature rise across the domain.
The corresponding reaction-rate distribution $\dot{\alpha}(x,t)$ indicates that polymerization occurs in a broad region rather than within a localized zone.
Moreover, the peak value of the reaction rate decreases progressively with time, reflecting the diminishing influence of boundary heating and indicating that the polymerization process eventually ceases beyond a certain distance.

In contrast, frontal polymerization proceeds without continuous external heat input.
The system is initialized with a prescribed temperature distribution $T(x,0) = T_m e^{-x/x_0}$, representing the localized energy input required to trigger the reaction, and the boundary condition is set to be adiabatic to eliminate any additional heat supply.
Due to the fast reaction kinetics, the heat released during polymerization diffuses into the neighboring region before dissipating, thereby triggering the reaction ahead of the front and enabling self-sustained propagation.
As a result, a sharp reaction front forms that separates the reacted and uncured regions and propagates through the material at a uniform velocity. Beyond this stable propagation mode, the system may also develop a pulsating front. In this mode, small temperature fluctuations near the front are amplified by the strong temperature dependence of the reaction rate. The front alternates between phases of rapid reaction and heating, during which the front temperature rises sharply, and phases in which little or no reaction occurs and the front cools. This cycle produces a periodic oscillation in both the front temperature and the propagation velocity. A detailed investigation of the stability of front propagation is presented in Section~\ref{sec:stability}.

The emergence of two distinct regimes from the same governing equations \eqref{eq:reac_diffusion} can be traced to differences in material properties. Although many of the thermal properties of the PDMS and DCPD systems are comparable, two kinetic parameters differ markedly and largely determine the contrasting behaviors observed in Fig.~\ref{fig:mode}. The first is the heat of reaction $H$, which in the DCPD system is nearly three times larger than in the PDMS system. This greater heat release strengthens the thermal feedback during curing and enables the reaction to sustain itself without external heating. The second key difference is the activation energy $E_a$, which is almost twice as large in the FP system. The higher activation energy makes the reaction rate much more sensitive to temperature, ensuring that the colder, unreacted region remains essentially inactive until the front arrives and raises the temperature sharply. These two factors together give rise to the localized, self-propagating front characteristic of frontal polymerization.

From the comparison above, we observe that frontal polymerization can cure the same amount of material in a much shorter time and with substantially lower energy input, highlighting its potential for efficient fabrication of polymer components. Therefore, the remainder of this manuscript focuses on applying the general polymerization theory to the case of frontal polymerization. Motivated by the sharp transitions and steep gradients near the front, we analyze the governing equation \eqref{eq:reac_diffusion} under the narrow reaction zone assumption, which enables a simplified and tractable analytical treatment of the propagation behavior.

\subsection{Narrow reaction zone assumption}
As observed in Fig.\ref{fig:mode}(b), the reaction zone is highly localized near the propagating front.
This observation motivates the introduction of the narrow reaction zone assumption, which has been widely used in combustion theory~\citep{zeldowitsch1988theory}.
The assumption states that the chemical reaction is considered to occur only at a single spatial point corresponding to the front position $x = \Phi(t)$.
Mathematically, this implies that the distribution of the reaction rate $\dot{\alpha}(x,t)$ can be approximated by a Dirac delta function concentrated at the front:
\begin{equation}\label{eq:G_delta}
\dot{\alpha}(x,t) = Q(t)\delta(x - \Phi(t)),
\end{equation}
where $Q(t)$ is the amplitude to be determined from the chemical kinetic relation~\eqref{chem_kinetic_1D}.

As a direct consequence of the narrow reaction zone assumption, away from the propagating front (i.e., $x \neq \Phi(t)$), no chemical reaction occurs, and thus the governing equations~\eqref{eq:reac_diffusion} reduce to:
\begin{equation}\label{eq:govern_reduce}
\left\{
\begin{aligned}
    c\dot{T} &= \kappa T_{xx} -h(T-T_0),\\
    \dot{\alpha} &= 0.\\
\end{aligned} 
\right.
\end{equation}

The influence of the localized reaction is incorporated through jump conditions for both the temperature field $T(x,t)$ and the degree of curing $\alpha(x,t)$ at the moving front location $x = \Phi(t)$.
The jump condition for $\alpha$ can be obtained directly by integrating $\dot{\alpha}$ given in \eqref{eq:G_delta} across an infinitesimal time interval enclosing the front:
\begin{equation}
    \llbracket\alpha \rrbracket  |_{x = \Phi(t)} = -\int_{t^-}^{t^+} \dot{\alpha}\mathrm{d}\tau = -\frac{Q(t)}{v(t)},
\end{equation} 
where $v(t) = \mathrm{d}\Phi(t)/\mathrm{d}t$ is the propagation speed of the moving front.
Here, the notation $\llbracket f(x) \rrbracket_{x_0} \equiv f(x_0^+) - f(x_0^-)$ denotes the jump of $f(x)$ across $x_0$.

Next, we consider the jump condition for the temperature field.
Since the governing equation~\eqref{eq:reac_diffusion} involves second-order spatial derivatives of temperature, discontinuities could arise in either the temperature $T$ itself or its spatial gradient $T_x$.
In the present analysis, we assume that the temperature field remains continuous across the front, while its gradient $T_x$  can exhibit a finite jump:
\begin{equation}
    \llbracket T_x\rrbracket|_{x = \Phi(t)} = \int_{\Phi(t)^-}^{\Phi(t)^+} T_{xx} dx= -\frac{\mathcal{F}\cdot Q(t)}{\kappa},
\end{equation}
where we have used the relation that $T_{xx} = -\mathcal{F}\dot{\alpha}/\kappa$ at the front location $x = \Phi(t)$ according to~\eqref{eq:reac_diffusion}, as well as the delta function approximation of $\dot{\alpha}$ given in \eqref{eq:G_delta}.

Finally, for analytical convenience, we introduce a coordinate system moving with the propagating front, defined as
\begin{equation}
\xi(x,t) = x - \Phi(t).
\end{equation}
The governing equation \eqref{eq:govern_reduce} can then be rewritten in terms of $T(\xi,t)$ and $\alpha(\xi,t)$ in this moving frame for $\xi\neq 0$:
\begin{equation}\label{eq:govern_front}
\left\{
\begin{aligned}
    c\left(\frac{\partial T}{\partial t} -v \frac{\partial T}{\partial \xi}\right)&= \kappa \frac{\partial^2 T}{\partial\xi^2} -h(T-T_0),\\
    \frac{\partial \alpha}{\partial t} -v \frac{\partial \alpha}{\partial \xi}&= 0,\\
\end{aligned} 
\right.
\end{equation}
accompanied by the two jump conditions at $\xi =0$:
\begin{equation}\label{eq:jump_front}
    \left.\left\llbracket\frac{\partial T}{\partial \xi}\right \rrbracket\right |_{\xi =0} = -\frac{\mathcal{F}\cdot{Q(t)}}{\kappa}, \quad \llbracket\alpha\rrbracket|_{\xi=0} = -\frac{Q(t)}{v(t)}.
\end{equation}

The governing equations~\eqref{eq:govern_front}, together with the jump conditions~\eqref{eq:jump_front}, describe the reaction–diffusion process associated with frontal polymerization characterized by a sharp propagating front.
At this stage, however, the reaction amplitude $Q(t)$ and the propagation velocity $v(t)$ remain undetermined and will be obtained from the chemical kinetic relation~\eqref{chem_kinetic_1D}.

\subsection{Traveling wave solution}
In Fig.~\ref{fig:mode}(b) we present the stable propagation mode of frontal polymerization, in which the temperature profile retains a fixed shape while translating at a constant speed.
This behavior can be described by a traveling-wave solution, where the temperature $T(\xi)$ and the degree of curing $\alpha(\xi)$ become time-independent in the moving frame attached to the front.
Consequently, the governing equations~\eqref{eq:govern_front} reduce to the following system of ordinary differential equations (ODEs) for $\xi \neq 0$:
\begin{equation}\label{eq:ode}
\left\{
\begin{aligned}
    &\kappa T^{\prime\prime}+c\cdot v T^{\prime}-h(T-T_0)=0,\\
    &\alpha^\prime =0,
\end{aligned}
\right.
\end{equation}
where the prime denotes derivative with respect to $\xi$. Note that both the amplitude $Q$ and the propagation velocity $v$ will be constant for the traveling wave solution. 

To simplify the analysis, we consider front propagation on an unbounded domain in the positive direction ($v >0$). This setup implies that the material ahead of the front (as $\xi \to +\infty$) remains uncured ($\alpha = 0$), while the material behind the front (as $\xi \to -\infty$) is fully cured ($\alpha = 1$). Besides, due to heat dissipation, the temperature far from the front (as $\xi \to \pm\infty$) is expected to converge to the environmental temperature $T_0$. Accordingly, we impose the following asymptotic boundary conditions:
\begin{equation}\label{eq:asymp}
    \lim_{\xi \rightarrow +\infty}\alpha(\xi) = 1,     \lim_{\xi \rightarrow -\infty}\alpha(\xi) = 0 ,\quad\quad     \lim_{\xi \rightarrow +\infty} T(\xi) = \lim_{\xi \rightarrow -\infty} T(\xi) = T_0
\end{equation}

For the degree of curing $\alpha(\xi)$, it follows from \eqref{eq:ode}$^2$ that $\alpha$ must be piecewise constant in the regions $\xi<0$ and $\xi>0$. Physically, this reflects the fact that curing occurs only at the front, while the material on either side remains either fully cured or entirely uncured. As a result, the traveling wave profile for $\alpha$ takes the form of a step function:
\begin{equation}\label{eq:alpha_distrubtion}
    \alpha(\xi) = \left\{
    \begin{aligned}
        &1\quad \xi <0,\\
        &0\quad \xi >0,\\
    \end{aligned}
    \right.
\end{equation}

For the temperature field $T(\xi)$, since \eqref{eq:ode}$^1$ is a homogeneous linear ODE with constant coefficients, it admits exponential functions as its general solution. Combined with the boundary condition \eqref{eq:asymp} and that $T(\xi)$ is continuous at $\xi =0$, we obtain the following form of  the traveling wave profile for $T$:
\begin{equation}\label{eq:temperature_stable}
    {T}(\xi) = \left\{
    \begin{aligned}
    &(T_m-T_0) e^{\frac{\xi}{s_1}}+T_0 \quad \xi<0,\\
    &(T_m-T_0) e^{-\frac{\xi}{s_2}}+T_0 \quad \xi>0,\\
    \end{aligned}
    \right.
\end{equation}
where $T_m$ is the front temperature to be determined and $s_1, s_2$ are characteristic lengths given by:
\begin{equation}\label{eq:s1s2}
\begin{aligned}
    s_1 = \frac{2\kappa}{-c\cdot v +\sqrt{c^2v^2+4\kappa h}},\quad
    s_2 = \frac{2\kappa}{c\cdot v +\sqrt{c^2v^2+4\kappa h}},
\end{aligned}
\end{equation}

At this stage, the traveling wave solution defined by \eqref{eq:alpha_distrubtion} and \eqref{eq:temperature_stable} contains two unknown parameters: the front temperature $T_m$ and the propagation velocity $v$. To determine them, we apply the two jump conditions associated with $\alpha(\xi)$ and $T(\xi)$ given in \eqref{eq:jump_front}, which introduce the amplitude $Q$  as an additional unknown. As a result, an extra relation is needed to fully determine the system.

While we prescribe the form of the reaction rate $\dot{\alpha}(\xi)$ in \eqref{eq:G_delta} as a Dirac delta function, its actual spatial distribution can be obtained from the chemical kinetic relation \eqref{chem_kinetic_1D} by substituting the known distribution for $T(\xi)$ and $\alpha(\xi)$ from \eqref{eq:temperature_stable} and \eqref{eq:alpha_distrubtion}, respectively. The amplitude $Q$ of the delta function, as defined in \eqref{eq:G_delta}, can then be approximated by integrating the kinetic relation across the spatial domain:
\begin{equation}
Q = \int_{-\infty}^{+\infty} \hat{\mathcal{G}}(\xi) \mathrm{d}\xi
= \int_{-\infty}^{+\infty} \mathcal{G}(T(\xi),\alpha(\xi), \lambda_e) \mathrm{d}\xi.
\end{equation}

By combining this expression for $Q$ with the jump conditions given in \eqref{eq:jump_front}, we obtain a closed system of nonlinear equations for the front temperature $T_m$ and the propagation velocity $v$:
\begin{equation}\label{eq:velocity}
    \begin{aligned}
         &T_m =T_0+\frac{\mathcal{F}}{c}\cdot\left(1+4\frac{\kappa h}{c^2v^2}\right)^{-1/2},\\
         &v  = C\cdot s_2(v) \cdot\frac{T_0}{T_*}\int_{T_*}^{T_m}e^{-\frac{E_a}{RT}}\frac{T-T_*}{T-T_0}\frac{1}{T} \mathrm{d} T,
    \end{aligned}
\end{equation}
where the expression for the characteristic length $s_2(v)$ is given in \eqref{eq:s1s2}.

By solving the coupled nonlinear equations for the front temperature 
$T_m$ and the propagation velocity $v$ in \eqref{eq:velocity}, the traveling wave solution of the frontal polymerization system is now fully determined.
Our analytical framework not only provides a closed-form prediction of the propagation velocity, but also enables investigation on how thermal dissipation and mechanical loading influences the propagation behavior through thermo-chemo-mechanical coupling.

\subsection{Propagation velocity}
Based on the analytical formula given in \eqref{eq:velocity}, this section focuses on investigating the influence of heat dissipation (characterized by the dissipation coefficient $h$) and mechanical loading (characterized by the elastic stretch $\lambda_e$) on the propagation dynamics characterized by the propagation velocity $v$. To isolate the effects of these two parameters, we first introduce a normalized velocity $\bar{v}$. Specifically, we define a reference velocity $v_0$ as the propagation speed in the absence of both heat dissipation ($h = 0$) and mechanical loading ($\lambda_e = 1$). Under these conditions, the front temperature reaches $T_m^0 = T_0 + H / c$, and the reference velocity $v_0$ is explicitly given by \eqref{eq:velocity}$^2$ as:
\begin{equation} \label{eq:reference_velocity}
v_0 = \sqrt{\frac{C\kappa}{c} \int_{T_0}^{T_m^0} e^{-E_a / RT} \frac{\mathrm{d}T}{T}}.
\end{equation}
Throughout this section, we focus on the normalized front velocity $\bar{v} = v/v_0$ as a means to characterize the relative change of propagation velocity.

\begin{figure} [h]
    \centering
    \includegraphics[width=1.0\columnwidth]{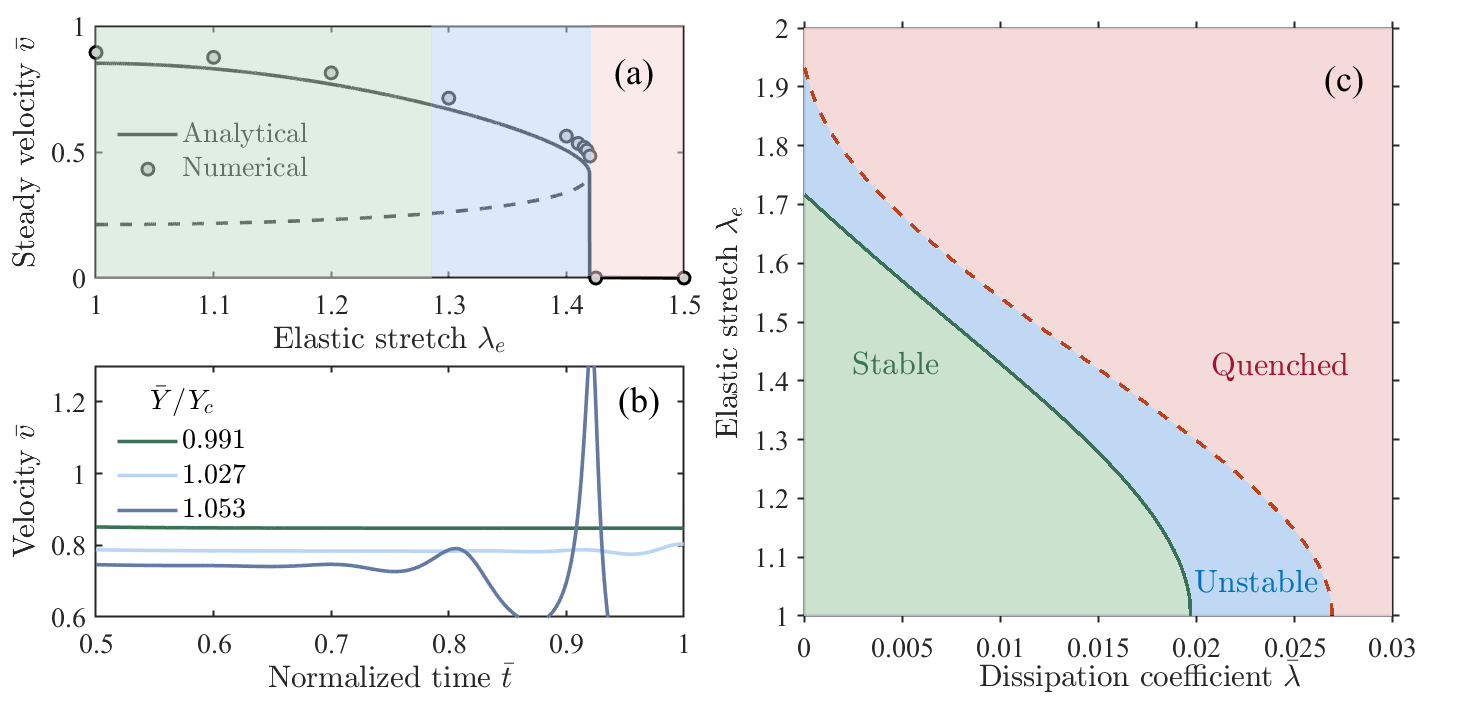}
    \caption{ Investigation of the influence of thermal dissipation and mechanical loading on propagation dynamics:
(a) Variation of normalized velocity $\bar{v} $ with the elastic stretch ${\lambda}_e$, evaluated with moderate dissipation $\bar{h} = 0.015$. The shaded regions indicate distinct propagation regimes, corresponding to the phase diagram in (c).
(b) Time evolution of the normalized front velocity $\bar{v}(\bar{t})$ from numerical simulations for different values of the normalized  parameter $\bar{Y}/Y_c$.
(c) Phase diagram showing the influence of the normalized dissipation coefficient $\bar{h}$ and elastic stretch $\lambda_e$ on the propagation regime.
}
    \label{fig:velocity}
\end{figure}

Figure~\ref{fig:velocity}(a) shows the normalized propagation velocity $\bar{v}$ as a function of the elastic stretch $\lambda_e$, with a normalized dissipation coefficient $\bar{h} \equiv \kappa h/(c v_0)^2 =0.015$.
The theoretical analysis reveals two branches of solutions below a critical value of $\lambda_e \sim 1.42$: an upper branch (solid line), where the velocity decreases gradually with increasing stretch, and a lower branch (dashed line), which exhibits the opposite trend.
Comparison between the analytical and numerical results indicates that only the traveling-wave solution associated with the upper branch is physically realized, and that the theoretical predictions agree well with the numerical simulations. The lower branch corresponds to an unstable mode, as will be demonstrated in the stability analysis of the traveling-wave solution in the next section. According to the stability analysis, the propagation becomes unstable in the blue region of Fig.~\ref{fig:velocity}(a), which is also confirmed by numerical simulations.
Before the onset of instability, a transient state is consistently observed in which the front velocity remains approximately steady for a short period before losing stability, as illustrated in Fig.~\ref{fig:velocity}(b).
The numerical velocity data points in Fig.~\ref{fig:velocity}(a) correspond to this temporally steady state.

As the elastic stretch $\lambda_e$ approaches a critical threshold, the two branches merge, and no solution exists beyond this point, indicating that the traveling-wave solution ceases to exist and the propagation is quenched ($v = 0$).
The abrupt drop in velocity is also observed by the numerical simulation results.
Most importantly, the sudden collapse of the front velocity beyond the critical stretch agrees with the experimental observation reported in our previous work~\citep{li2024mechanical}, where a binary response was observed: the front velocity remained nearly constant over a wide range of applied loads but dropped abruptly once a critical level was exceeded.

Similarly, the same trend is observed for the case when the elastic stretch is fixed and the dissipation coefficient is varied.
To examine the combined influence of heat dissipation and mechanical loading, Fig.~\ref{fig:velocity}(c) presents a phase diagram where the transition to the quenched state is indicated by the dashed line.
The phase diagram shows that both heat dissipation and mechanical loading can induce quenching, and their effects can be additive.
For example, at higher dissipation levels, quenching occurs at much lower mechanical loading.

\subsection{Stability analysis}
\label{sec:stability}
\noindent In the previous sections, we derived a traveling-wave solution for frontal polymerization, which corresponds to the most commonly observed mode of stable front propagation. However, experimental studies have also reported regimes in which the propagation loses stability and exhibits oscillatory or pulsating behavior rather than maintaining a steady profile \citep{pojman1995spin, lloyd2021spontaneous}. Similar phenomena have been extensively studied in the context of flame propagation, where the Zeldovich number \citep{zeldovich1985mathematical} has been proposed as a parameter that characterizes the onset of thermal instability:
\begin{equation}\label{eq:zeldovich}
Z = \frac{E_a}{R T_m^2} (T_m - T_0).
\end{equation}
Perturbation analyses \citep{matkowsky1978propagation} show that the propagation is stable when $Z < Z_c = 4 + 2\sqrt{5} \approx 8.4$ and becomes unstable for $Z > Z_c$. While this criterion describes thermal instability in adiabatic systems, it does not incorporate heat loss or mechanical loading, which are essential in frontal polymerization.

In this section, we perform a linear stability analysis of the traveling-wave solution derived earlier and introduce a new dimensionless criterion for the onset of instability that incorporates the effects of both heat loss and mechanical loading. For simplicity, we restrict attention to perturbations in the temperature field and assume that the degree of curing $\alpha$ remains fixed at its traveling-wave profile given in \eqref{eq:alpha_distrubtion}. We introduce an infinitesimal perturbation $\delta T(\xi,t)$ to the traveling-wave temperature profile $T(\xi)$ obtained in \eqref{eq:temperature_stable}, and express the perturbed temperature field $\tilde{T}(\xi,t)$ as
\begin{equation}
\tilde{T}(\xi,t) = T(\xi) + \delta T(\xi,t).
\end{equation}

Since the propagation velocity $v$ appears explicitly in the governing equation \eqref{eq:govern_front}, we must also account for its perturbation induced by the perturbed temperature field. Similar to the perturbed temperature field $\tilde{T}(\xi,t)$, we thus write the perturbed velocity $\tilde{v}(t)$ as:
\begin{equation}
\tilde{v}(t) = {v} + \delta v(t),
\end{equation}
where $v$ is the velocity for the traveling wave solution.

Substituting the perturbed fields into the governing equation \eqref{eq:govern_front} and linearizing around the base solution ${T}(\xi)$ yields the evolution equation for the temperature perturbation $\delta T(\xi,t)$:
\begin{equation}\label{eq:pde_perturbation}
\frac{\partial (\delta T)}{\partial t}
- \frac{\kappa}{c}  (\delta T)^{\prime\prime}
- {v}  (\delta T)^{\prime}
+ \frac{h}{c}  \delta T
= \delta v \cdot {T}^{\prime},
\end{equation}
where ${T}^{\prime} (\xi)$ is the spatial derivative of the base solution and can be obtained from \eqref{eq:temperature_stable}. 

To close the system, we need to establish a connection between the velocity perturbation $\delta v$ and the temperature perturbation $\delta T(\xi,t)$. Since the propagation velocity $v$ is explicitly determined by the front temperature $T_m$, as given in \eqref{eq:velocity}, it is natural to assume that $\delta v$ is directly influenced by the perturbation in temperature at the front location $\xi = 0$ throught the following linear relation:
\begin{equation}
\delta v(t) = Y \cdot \delta T(0,t),
\end{equation}
where $Y$ is a constant that characterizes the sensitivity of the front velocity to fluctuations in the front temperature. Physically, $Y$ quantifies the strength of the temperature dependence in the propagation dynamics: a larger value of $Y$ implies that small variations in temperature at the front lead to more significant changes in the front velocity. The value of $Y$ can be evaluated by differentiating the expression for the front velocity with respect to $T_m$ in \eqref{eq:velocity}, yielding:
\begin{equation}\label{eq:Y}
    Y  \overset{\text{def}}{=} \frac{\delta v}{\delta T_m} = Cs_2\frac{T_0}{T_*} e^{-\frac{E_a}{RT_m}}\cdot\frac{T_m-T_*}{T_m-T_0}\cdot\frac{1}{T_m}\cdot\left(1-\frac{\mathrm{d}s_2}{\mathrm{d}v}\cdot\frac{v}{s_2}\right)^{-1}.
\end{equation}

Inspired by the dimensionless Zeldovich number, we introduce a dimensionless form of the sensitivity parameter $Y$ as:
\begin{equation}\label{eq:Y_bar}
    \bar{Y} = \frac{2\mathcal{F}}{cv} Y.
\end{equation}
Under the adiabatic and stress-free condition ($h = 0$, $\lambda_e = 0$), it can be shown that this quantity reduces to the classical Zeldovich number, and one obtains the relation $\bar{Y} = Z$. The detailed derivation is provided in \ref{app:Zeldovich}.

The governing equation \eqref{eq:pde_perturbation} is linear with time-independent coefficients, which allows us to analyze it using a spectral decomposition approach. We therefore assume a normal mode solution with exponential time dependence, characterized by a spectral parameter $\omega$:
\begin{equation}\label{eq:normal_mode}
    \delta T(\xi,t) = f(\xi) e^{\omega t},
\end{equation}
where $f(\xi)$ describes the spatial structure of the perturbation, and $\omega$ governs its temporal evolution. Note that here $\omega$ is in general a complex number: the real part determines whether the perturbation grows or decays over time, while the imaginary part corresponds to oscillatory behavior.
 Substituting this ansatz into \eqref{eq:pde_perturbation} transforms the problem to an ODE for $f(\xi)$, parameterized by $\omega$:
\begin{equation}\label{eq:ode_f}
    \frac{\kappa}{c} f^{\prime\prime} + {v} f^\prime - \left(\frac{h}{c} + \omega\right)f = - Y\cdot f(0)\cdot{T}^\prime.
\end{equation}

Following the asymptotic boundary conditions for the temperature field given in \eqref{eq:asymp}, it is reasonable to assume that the spectral component of the perturbation also vanishes at remote field:
\begin{equation}\label{eq:asymp_f}
    \lim_{\xi \to +\infty} f(\xi) = \lim_{\xi \to -\infty} f(\xi) = 0.
\end{equation}

Combining the governing equation \eqref{eq:ode_f} with the boundary conditions \eqref{eq:asymp_f}, we can solve explicitly for $f(\xi)$:
\begin{equation}\label{eq:solution_f}
    \frac{f(\xi)}{f(0)} = \left\{
    \begin{aligned}
    &\frac{Y ({T}_m-T_0)}{\omega s_1} e^{\frac{\xi}{s_1}} + \left(1-\frac{Y ({T}_m-T_0)}{\omega s_1}\right)e^{\frac{\xi}{p_1}} \quad \xi<0,\\
    &-\frac{Y ({T}_m-T_0)}{\omega s_2} e^{-\frac{\xi}{s_2}} + \left(1+\frac{Y ({T}_m-T_0)}{\omega s_2}\right)e^{-\frac{\xi}{p_2}}  \quad \xi>0,\\
    \end{aligned}
    \right.
\end{equation}
where ${T}_m$ is the front temperature of the base solution and $p_1, p_2$ are characteristic lengths given by:
\begin{equation}\label{eq:p1p2}
\begin{aligned}
    p_1 = \frac{2\kappa}{-c\cdot {v} +\sqrt{c^2{v}^2+4\kappa(h + c\omega)}},\quad
    p_2 = \frac{2\kappa}{c\cdot {v} +\sqrt{c^2{v}^2+4\kappa(h + c\omega)}}.
\end{aligned}
\end{equation}

Similar to the base solution given in \eqref{eq:temperature_stable}, the perturbation of the temperature field also exhibits a discontinuity in its spatial derivative at the front $\xi = 0$. A jump condition for the perturbation field can be derived by linearizing  the jump condition of temperature field \eqref{eq:jump_front} around the base solution, which gives:
\begin{equation}\label{eq:jump_perturbation}
[f^\prime]|_{\xi=0} = -\frac{\mathcal{F}}{\kappa} \cdot Y \cdot f(0).
\end{equation}


Substituting the solution for the perturbation field $f(\xi)$ given in \eqref{eq:solution_f} into the jump condition \eqref{eq:jump_perturbation}, we obtain a characteristic equation for the spectral parameter $\omega$, which turns out to be a quadratic equation in form. Solving this equation yields the following explicit expression for $\omega$:
\begin{equation}\label{eq:omega}
\omega = \frac{cv^2}{8\kappa} \left( \frac{\bar{Y}^2}{4} - \frac{2}{\theta} \bar{Y} - \theta^2 \pm (\frac{\bar{Y}}{2} - \theta) \sqrt{(\frac{\bar{Y}}{2} + \theta)^2 - \frac{4}{\theta} \bar{Y}} \right),
\end{equation}
where  we have introduced a dimensionless quantity $\theta$ to simplify the expression:
\begin{equation}
\theta = \sqrt{1 + \frac{4\kappa h}{c^2 v^2}},
\end{equation}
which quantifies the influence of heat loss. In the absence of heat loss ($h = 0$), we recover the limiting case $\theta = 1$.

The stability of the traveling wave solution is determined by the sign of the real part of $\omega$. Specifically, the system is stable if the perturbations decay over time, i.e.,
\begin{equation}
\mathrm{Re}(\omega) < 0.
\end{equation}

Since the expression for $\omega$ in \eqref{eq:omega} contains two branches (due to the $\pm$ term), stability requires that both branches of the solution yield negative real parts. This ensures that no mode leads to unbounded temporal growth.
The resulting condition on $\bar{Y}$ can be expressed as the following critical threshold:
\begin{equation}\label{eq:stability}
\bar{Y} < Y_c = \left\{
\begin{aligned}
&\frac{4}{\theta} + 2\sqrt{\theta^2 + \frac{4}{\theta^2}} \quad\quad & 1 \le \theta \le \frac{\sqrt{6}}{2}, \\
&\frac{2\theta^3}{\theta^2 - 1} \quad\quad\quad\quad\quad & \theta > \frac{\sqrt{6}}{2}.
\end{aligned}
\right.
\end{equation}
In the limiting case without heat loss ($h = 0$), we have $\theta = 1$, and the critical value reduces to $Y_c = 4 + 2\sqrt{5}$. From the derivation in \ref{app:Zeldovich}, we have $Y =  Z$ under adiabatic, stress-free conditions. Substituting this relation into the stability condition $Y < Y_c$ gives $Z < Z_c = 4+ 2\sqrt{5}$, which coincides with the classical prediction from Zeldovich’s theory for flame propagation \citep{zeldovich1985mathematical}. Thus, our criterion recovers the classical result in the adiabatic limit and generalizes the stability condition to frontal polymerization in the presence of heat loss and mechanical loading.

Equation~\eqref{eq:stability} serves as the stability criterion for the traveling-wave solution derived in the previous sections.
As a validation of this criterion, Fig.~\ref{fig:velocity}(b) shows the evolution of the front velocity obtained from numerical simulations, where a clear transition marking the loss of stability is observed when $\bar{Y} > Y_c$.
Furthermore, in the phase diagram shown in Fig.~\ref{fig:velocity}(c), the solid line separating the stable and unstable regions is determined directly from the analytical stability criterion~\eqref{eq:stability}.
The phase diagram shows that, during the transition from a stable traveling-wave solution to the quenched state, the system first enters an unstable regime before propagation is quenched.
It also demonstrates that the stability of front propagation can be influenced by mechanical loading, indicating that local stress can either promote or suppress stable propagation depending on the loading conditions.

\section{Coupling mechanism II: Force response}
\noindent
In the previous section we investigated the first coupling mechanism in the polymerization process, namely the influence of the local stress state on the reaction kinetics, with particular focus on the case of frontal polymerization. In this section we turn to the complementary coupling mechanism, which concerns how the stress evolves during the polymerization process. Residual stress accumulation is a well-known issue in polymerization, as the volume mismatch between the polymer and surrounding materials such as molds or reinforcing fibers can induce significant stress during curing. In frontal polymerization this effect is further amplified by the sharp temperature and curing gradients across the propagating front. As a result, understanding how stress develops and accumulates during FP is essential for predicting potential warping and ensuring structural integrity in cured components.

Here, we adopt the same one-dimensional simplification used in the previous section. To isolate the effect of front propagation on the mechanical response, we assume that the front advances at a constant velocity corresponding to the stable traveling-wave regime.
Specifically, we theoretically investigate the evolution of the axial force in a uniaxial frontal polymerization setup under a fixed prescribed elongation $\Delta$ (displacement control), as illustrated in Fig.~\ref{fig:uniaxial}(a).
The theoretical predictions obtained from this analysis are then qualitatively compared with experimental results to further assess and validate the proposed model.

\subsection{Experimental setup $\&$ key observations}
\begin{figure} [h]
    \centering
    \includegraphics[width=0.8\columnwidth]{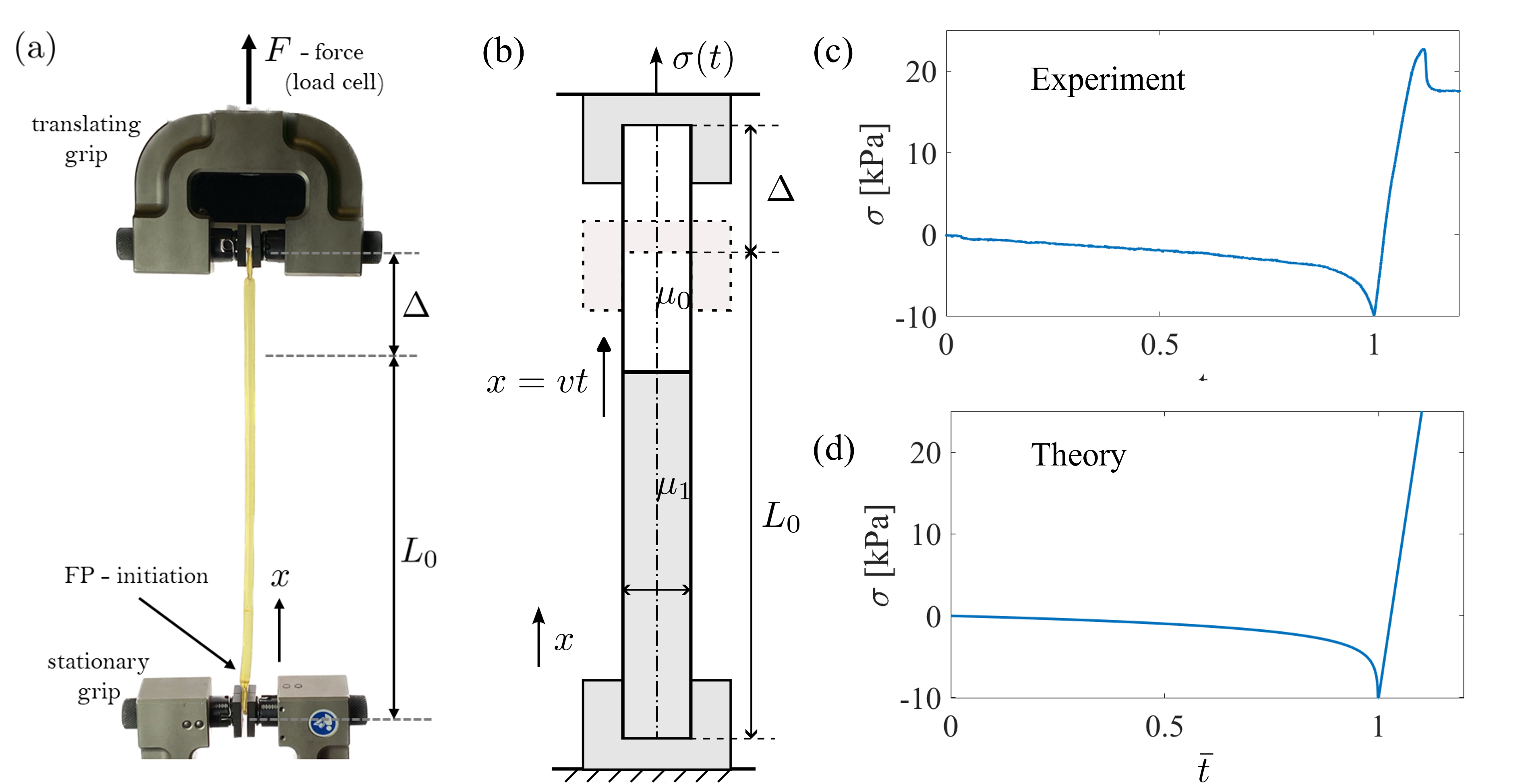}
    \caption{(a) Illustration of the experimental set up. A soft polymer sample is mounted on the mechanical testing machine that controls its length and measures the applied force; (b) Simplified schematic of the uniaxial setup, the front propagation behavior is simplified as a shock-like wave leading to discontinuities in distribution of material properties; (c) Experimental measurement of the uniaxial stress as a function of time, here for the convenience of comparison, the time is normalized as $\bar{t} =t/t_0$, where $t_0 = 1300s$ denotes the time at which the sample is fully cured. Thus $\bar{t}=1$ represents the instant when the front reach the end of the sample. (d) Theoretical prediction of the stress response during FP process.}
    \label{fig:uniaxial}
\end{figure}

\noindent
The soft gel-like solid specimen, with an initial length $L_0 = 20\ \mathrm{cm}$, was mounted on an Instron  machine, as shown in Fig.~\ref{fig:uniaxial}(a). Prior to initiating the FP process, the top grip was displaced by a fixed amount $\Delta$, and the system was allowed to relax long enough to eliminate any rate-dependent viscoelastic effects. The FP process was then triggered at the bottom end of the sample using a soldering iron, initiating an upward-propagating front while the applied displacement $\Delta$ was held constant. Throughout the experiment, the axial force response $F(t)$ was continuously recorded by the Instron load cell, and the propagation of the front was tracked using an infrared camera.

Experimental observations show that the polymerization front travels upward at an approximately constant speed. The measured force response, presented in Fig.~\ref{fig:uniaxial}(c), exhibits a distinct multi-stage behavior. As the front propagates, a compressive force gradually builds up. Once the front reaches the top and the sample becomes fully polymerized, a sharp transition occurs, followed by stabilization of the force at a positive (tensile) value after cooling.
This final tensile state points to a permanent evolution of the stress-free configuration during the polymerization process. These observations highlight the necessity of introducing the concept of transformation deformation and its evolution law in order to accurately model the force response and account for residual stress accumulation, which is a central focus of the following analysis. 

\subsection{Analytical formulation}
\noindent
A simplified schematic of the uniaxial experimental setup for frontal polymerization is shown in Fig.~\ref{fig:uniaxial}(b). To enable analytical treatment, we begin by introducing several simplifying assumptions. Motivated by the step-function distribution of $\alpha$ obtained from the traveling-wave solution in \eqref{eq:alpha_distrubtion}, we assume that the degree of curing has the spatial distribution:
\begin{equation}
\begin{aligned}
    \alpha(x,t) = \left\{
\begin{aligned}
     &1\quad x< vt,\\
&0\quad x> vt,\\
\end{aligned}    
\right.
\end{aligned}
\end{equation}
where $v$ is the constant front speed. This step-function representation implies that the reaction is confined to an infinitesimally thin region at the front, and therefore the reaction rate takes the form of a Dirac delta distribution, consistent with the narrow reaction-zone assumption used earlier:
\begin{equation}\label{delta}
    \dot{\alpha} = v\cdot\delta(x-vt).
\end{equation}

Next, we prescribe the distribution of the temperature field. Although the full temperature profile is given in \eqref{eq:temperature_stable}, we further simplify the analysis by assuming that the time scales associated with thermal diffusion and heat exchange with the environment are much longer than that of the chemical reaction. Under this assumption, the temperature evolution is governed entirely by the exothermic heat released during polymerization. Consequently, the temperature can be expressed directly in terms of the degree of curing as
\begin{equation} T(\alpha) = T_0 + (T_m - T_0) \cdot \alpha, \end{equation} where $T_0$ is the initial (ambient) temperature, and $T_m$ denotes the maximum temperature reached when $\alpha = 1$. This assumption amounts to neglecting both heat loss and thermal diffusion, while retaining the essential structure of the temperature profile in \eqref{eq:temperature_stable}.

With this relation between temperature and degree of curing, the volume change relation given in \eqref{volume} leads to the following distribution:
\begin{equation}\label{volume_distrubtion}
J_i(x,t) = 1 + \left[\beta(T_m - T_0) - \gamma\right]\alpha(x,t) = \left\{
\begin{aligned}
     &J_0\quad x< vt,\\
&1\quad x> vt,\\
\end{aligned}    
\right. 
\end{equation} 
where $J_0= 1+\beta(T_m - T_0) - \gamma$ is a constant summarizing the combined effects of thermal expansion and chemical shrinkage. Accordingly, the volumetric deformation gradient can be expressed as $\bm{F}_v = J_i^{1/3}\bm{1}$.

In the uniaxial experimental setup, considering the incompressibility condition \eqref{incompressible_condition} and the symmetric geometry within the cross section, the elastic deformation gradient $\bm{F}_e$ and the transformation deformation gradient $\bm{F}_*$ should both have the following structure:
\begin{equation}
\label{uniaxial_stretch}
    \bm{F}_e = \begin{bmatrix}
        \lambda_e&0&0\\
        0&\lambda_e^{-1/2}&0\\
        0&0&\lambda_e^{-1/2}\\
    \end{bmatrix},
    \bm{F}_* = \begin{bmatrix}
        \lambda_*&0&0\\
        0&\lambda_*^{-1/2}&0\\
        0&0&\lambda_*^{-1/2}\\
    \end{bmatrix},
\end{equation}
where $\lambda_e(x,t)$ and $\lambda_*(x,t)$ denote the distribution of elastic and transformation stretches, respectively, in the axial direction.

Based on the proposed structure of the elastic deformation gradient and the constitutive relation \eqref{const}$^1$, along with the traction-free condition on the lateral surface of the cylindrical sample, the only non-zero component of the Cauchy stress tensor is the axial component. We denote this as $\bm{T}_{11}(x,t)= \sigma(x,t) $, representing the axial stress distributed along the sample. It is related to the elastic stretch $\lambda_e$ through: 
\begin{equation}\label{stress_strain}
    \sigma(x,t) = \left\{
\begin{aligned}
     &\mu_1(\lambda_e^2-\lambda_e^{-1}),\quad x< vt\\
&\mu_0(\lambda_e^2-\lambda_e^{-1}),\quad x> vt\\
\end{aligned}    
\right.
\end{equation}
where $\mu_0$ and $\mu_1$ denote the shear modulus before and after polymerization, respectively.

With the given structure of the Cauchy stress, the momentum balance equation \eqref{mechanical equilibrium} reduces to: 
\begin{equation}
\frac{\partial \sigma(x,t)}{\partial x} = 0. \end{equation} 
This equilibrium condition indicates that the axial stress $\sigma$ is spatially uniform, which allows us to simplify the notation by writing $\sigma(t)$. This uniformity of axial stress makes it a natural and convenient choice to characterize related field variables. In particular, we can invert the stress–stretch relation \eqref{stress_strain} to express the elastic stretch $\lambda_e$ in terms of the axial stress $\sigma$:
\begin{equation}\label{strain_stress}
    \lambda_e(x,t) = \left\{
\begin{aligned}
     &s(\sigma(t)/\mu_1)\quad x< vt\\
&s(\sigma(t)/\mu_0)\quad x> vt\\
\end{aligned}    
\right.
\end{equation}
where $s(\cdot)$ is a known compliance function defined implicitly by inverting $\sigma/\mu = \lambda_e^2 - \lambda_e^{-1}$. This distribution indicates that, while the stress field remains continuous across the front, the elastic stretch exhibits a jump due to the abrupt change in the elastic modulus upon curing.

To complete the kinematic description, we now determine the evolution of the transformation stretch $\lambda_*(x,t)$. In the uniaxial setting, the general flow rule \eqref{flow_rule} in tensor form reduces to an evolution equation of $\lambda_*$. By substituting the uniaxial forms of the deformation gradients \eqref{uniaxial_stretch} into the general formulation, we obtain:
\begin{equation}\label{ode_lambda}
\left(\frac{2}{3}\lambda_e^2+\frac{1}{3}\lambda_e^{-1}\right)\frac{\dot{\lambda_*}}{\lambda_*} = \frac{\mu_H}{3\mu^2} \sigma \dot{\alpha},
\end{equation}
which relates the evolution of $\lambda_*(x,t)$ with the axial stress $\sigma$ and the reaction rate $\dot{\alpha}$. As shown in \eqref{delta}, $\dot{\alpha}(x,t)$ is a Dirac delta function concentrated at the propagating front ($x= vt$). Integrating \eqref{ode_lambda} with the initial condition $\lambda_*(x,0) = 1$ yields the following analytical expression for the transformation stretch:
\begin{equation}\label{transformation_stretch}
    \lambda_*(x,t) = \left\{
\begin{aligned}
     &\frac{s(\sigma(x/v)/\mu_0)}{s(\sigma(x/v)/\mu_1)}\quad x< vt\\
&1\quad \quad \quad \quad \quad \quad  x> vt\\
\end{aligned}    
\right.
\end{equation}

Two key observations arise from the analytical expression of the transformation stretch \eqref{transformation_stretch}. First, $\lambda_*(x,t)$ depends on the axial stress at the moment the front passes ($t_0 = x/v$), rather than the current time. This implies that the transformation stretch is determined by the stress history experienced at the moment of curing and thus inherently encodes the loading path. Second, comparing with the elastic stretch–stress relation \eqref{strain_stress}, we observe that $\lambda_*$ at a given point corresponds to the ratio between the elastic stretches immediately ahead of and behind the front as it crosses that point. This observation provides a practical way to determine the transformation deformation from the jump of elastic deformation across the front.

Now we have established the analytical expression of the distribution of the volume change \eqref{volume_distrubtion}, elastic stretch \eqref{strain_stress}, the transformation stretch \eqref{transformation_stretch}, we can now evaluate the total axial stretch $\lambda$ which is directly linked to the observed deformation, follows from the kinematic decomposition relation \eqref{decomp_total}:
\begin{equation}\label{lambda}
    \lambda(x,t) \equiv \frac{dx}{dX}= J_i^{1/3}\lambda_e\lambda_*= \left\{
\begin{aligned}
     &J_0^{1/3} s(\sigma(t)/\mu_1)\frac{s(\sigma(x/v)/\mu_0)}{s(\sigma(x/v)/\mu_1)}\quad x< vt\\
&s(\sigma(t)/\mu_0)\quad\quad\quad\quad\quad\quad\quad\quad x> vt\\
\end{aligned}    
\right.
\end{equation}

The total displacement of the sample, denoted by $\Delta(t)$, can now be computed based on the spatial distribution of the total axial stretch $\lambda(x,t)$. Since $\lambda$ is expressed in terms of the deformed spatial coordinate $x$ rather than the reference coordinate $X$, we relate the total displacement to the stretch as following: 
\begin{equation}\label{disp_def} 
\Delta(t) = \int_0^{L_0} (\lambda - 1)  dX = \int_0^{l(t)} \frac{\lambda(x,t) - 1}{\lambda(x,t)} dx, 
\end{equation}
where $l(t) = L_0 + \Delta(t)$ is the deformed length of the specimen at time $t$. Substituting the explicit form of $\lambda(x,t)$ in \eqref{lambda} into \eqref{disp_def} and performing the integration yields a theoretical relation between the applied displacement and the stress response of the sample during the frontal polymerization process. 

To investigate the force response under displacement-controlled conditions, we now impose the boundary condition that the total displacement remains fixed throughout the polymerization process, i.e., $\Delta(t) = \mathrm{const}$. In differential form, this constraint reads $\dot{\Delta}(t) = 0$. Substituting the expression for $\Delta(t)$ from \eqref{disp_def}, this condition translates into a differential equation governing the evolution of the axial stress $\sigma(t)$:
\begin{equation}\label{sigma_govern_1}
    \dot{\sigma}(t) = -v(1-J_0^{1/3})\frac{s(\sigma(t)/\mu_0)}{s^\prime(\sigma(t)/\mu_0)}\cdot\left(\frac{l-vt}{\mu_0}+J_0^{-1/3}\frac{v}{\mu_1}\int_0^t\frac{s(\sigma(\tau)/\mu_1)}{s(\sigma(\tau)/\mu_0)}d\tau\right)^{-1}.
\end{equation}

This equation characterizes the evolution of axial stress as the front propagates under fixed displacement. Notably, if the volume change effect is absent (i.e., $J_0 = 1$), then $\dot{\sigma}(t) = 0$, indicating that the stress will remain constant throughout the process. This observation aligns with the consistency condition stated in \eqref{eq:consistent}, which asserts that no additional macroscopic deformation implies no change in stress.

The governing equation \eqref{sigma_govern_1} applies during the propagation stage, where $0 < t < l/v$, as the polymerization front moves through the sample. Once the front reaches the end of the specimen ($t = l/v$), the subsequent mechanical response is driven by the recovery of the volume change ratio $J_i$ due to cooling. For analytical simplicity, we assume that the sample cools uniformly back to room temperature following Newton's law of cooling. Under this assumption, the volume change along the axis evolves during the cooling stage ($t>l/v$) as: 
\begin{equation} \label{volume_evolve}
J_i(t) = J_1 +(J_0-J_1) e^{-k(t-l/v)},
\end{equation}
where $J_1 = 1 - \gamma$ is the volume change ratio resulting solely from chemical shrinkage after the temperature returns to the ambient value $T_0$, and $k$ is the cooling rate constant.

Substituting the time-dependent volume change ratio $J_i(t)$ from \eqref{volume_evolve} into the expression of total displacement \eqref{disp_def}, and applying the constraint $\dot{\Delta}(t) = 0$, we obtain a differential equation governing the evolution of the axial stress $\sigma(t)$ during the cooling stage ($t > l/v$):
\begin{equation}\label{sigma_govern_2} \dot{\sigma}(t) = -\frac{\mu_1}{3}\frac{s(\sigma(t)/\mu_1)}{s^\prime(\sigma(t)/\mu_1)}\cdot\frac{\dot{J}_i(t)}{J_i(t)}. \end{equation}

Accordingly, the complete evolution of the axial stress $\sigma(t)$ is described by two regimes: the propagation stage ($0 < t < l/v$), governed by \eqref{sigma_govern_1}, and the cooling stage ($t > l/v$), governed by \eqref{sigma_govern_2}.

\subsection{Comparison with Experimental Measurements}

\noindent To evaluate the predictive capability of the theoretical framework, we compare the analytically derived force response with experimental measurements, as shown in Fig.~\ref{fig:uniaxial}(c,d). Subplot (c) presents the experimentally recorded axial stress $\sigma(t)$, while subplot (d) displays the corresponding theoretical prediction obtained from solving the governing equations \eqref{sigma_govern_1} and \eqref{sigma_govern_2}. The material and kinetic parameters used in the theoretical analysis are chosen to match the experimental conditions. Specifically, the shear modulus in the gel state is taken as $\mu_0 = 140~\mathrm{kPa}$, the chemical hardening modulus as $\mu_H = 1.4~\mathrm{GPa}$, and the initial volumetric Jacobian as $J_0 = 1.01$. The front propagation velocity is prescribed as $v = 0.92~\mathrm{cm/min}$, and the cooling rate constant is set to $k = 1.92\times10^{-5}~\mathrm{s^{-1}}$.

Qualitatively, the theory captures the key features of the experimental response: the initial development of compressive stress as the front propagates, followed by a sharp reversal and buildup of tensile stress after full polymerization. This agreement supports the validity of the proposed evolution law for the transformation deformation and its role in residual stress generation.

It is worth noting that in the experiment, the final rise in tensile force eventually plateaus. This plateau is likely due to slippage at the grips, which limits further stress buildup. In contrast, the theoretical model, absent of such failure mechanisms, predicts a continuously increasing tensile stress post-polymerization, potentially reaching significantly higher values. This divergence underscores the importance of accounting for residual stress when designing structures involving FP, as it can lead to substantial internal loads if not properly constrained.

\section{Conclusion}
\noindent Polymerization processes inherently involve the interplay of thermodynamics, chemical kinetics, and mechanical response. As a result, the ability to model and predict such complex coupling effects is essential for the industrial manufacturing of polymer components, particularly in large-scale fabrication and geometrically complex systems. Frontal polymerization (FP) has recently emerged as a promising alternative to conventional curing methods, offering rapid, energy-efficient processing that makes it especially suitable for large-scale applications. However, FP features a propagating reaction front across which material properties, such as stiffness, temperature, and conversion, change abruptly, amplifying the coupling effects and necessitating further investigation. Prior work has highlighted different aspects of these interactions: for instance, \citet{kumar2022surface, wijaya2025thermo} and our previous study \citep{li2024mechanical} investigated the evolution of the stress-free configuration during FP. Additionally, our recent work \citep{li2024mechanical} also provided the first experimental observation that local stress can influence chemical kinetics. These observations suggest a two-way coupling in FP between mechanical stress and reaction dynamics. Therefore, a unified, thermodynamically consistent framework capable of capturing coupling mechanisms in FP, or more broadly in polymerization processes involving large deformations, is critically needed.

In this work, we formulated a fully coupled, large-deformation thermo-chemo-mechanical framework for polymerization process. The constitutive model is derived from the second law of thermodynamics by constructing a free energy imbalance, from which a thermodynamic driving force for the polymerization reaction is defined. We proposed a chemical kinetic relation based on this driving force to ensure thermodynamic consistency. To capture the evolution of the stress-free configuration, we introduced a multiplicative decomposition of the deformation gradient and defined a transformation deformation gradient whose evolution follows a rate-from flow rule inspired by plasticity theory. This general formulation provides a consistent foundation for analyzing polymerization processes with strong mechanical coupling effects.

We then specialized the framework to frontal polymerization and analyzed the propagation behavior using a reduced one dimensional model together with a narrow reaction zone approximation. Closed-form expressions were obtained for the propagation velocity, revealing how both heat loss and mechanical loading influence the propagation dynamics. In particular, the theory predicts a sudden drop to zero in the propagation velocity once the applied mechanical loading exceeds a critical threshold, explaining the binary response observed in our previous experiments \citep{li2024mechanical}. In addition, the analysis yields a stability criterion that governs the existence of steady traveling-wave solutions. This criterion generalizes the classical Zeldovich condition by incorporating mechanical effects, and reduces to the classical Zeldovich number in the adiabatic, stress-free limit. Based on this criterion, we constructed a phase diagram that delineates regimes of stable propagation, instability, and quenching.
We also examined the mechanical response during FP by analyzing the axial force evolution in a uniaxial setup under displacement control. By modeling the reaction front as a shock like transition in material properties, we derived analytical predictions for the stress development as the front propagates. The theoretical results show good qualitative agreement with experiments and provide insight into the mechanisms that generate and accumulate residual stress during polymerization.

Overall, this study provides a unified perspective on the coupled propagation and mechanical response in polymerization processes. The framework offers guidance for controlling propagation behavior and mitigating residual stresses in FP based manufacturing. Future work will focus on extending the analysis to more complex loading conditions and geometries and on developing numerical implementations for three dimensional simulations of the fully coupled theory.

\section*{Acknowledgments} 
The authors would like to thank Professor John Pojman (LSU) and Dominic Adrewie (LSU) for the helpful discussions. 

\appendix
\section{Constitutive relation of stress with incompressibility constraint}\label{app:incomp}
\noindent The first term of \eqref{major} can be rewritten as:
\begin{equation}
\label{tptp}
\begin{aligned}
    &\left(\rho\frac{\partial \psi_r}{\partial \bm{C}_e}-\frac{1}{2}\bm{S}\right):\dot{\bm{C}}_e
    =\bm{F}_e\left(\rho\frac{\partial \psi_r}{\partial \bm{C}_e}-\frac{1}{2}\bm{S}\right)\bm{F}_e^T:{\bm{D}}_e,
\end{aligned}
\end{equation}
where we have used the fact that both $\frac{\partial \psi_r}{\partial \bm{C}_e}$ and $\bm{S}$ are symmetric tensors.

The incompressible constraint on the elastic deformation expressed in \eqref{incompressible_condition} implies that $\bm{D}_e$ is a deviatoric tensor. 
Since elastic deformation should not contribute to the energy dissipation, the first term in \eqref{major} should vanish for any deviatoric $\bm{D}_e$, which means the first factor in \eqref{tptp} has to be a spherical tensor:
\begin{equation}
\begin{aligned}
    \bm{F}_e\left(\rho\frac{\partial \psi_r}{\partial \bm{C}_e}-\frac{1}{2}\bm{S}\right)\bm{F}_e^T = q\bm{1},
\end{aligned}
\end{equation}
where $q$ is an arbitrary scalar field.

Using \eqref{second_piola}, the Cauchy stress $\bm{T}$ can be expressed as:
\begin{equation}
\begin{aligned}
    \bm{T} &= 2\rho\bm{F}_e\frac{\rho\partial \psi_r}{\partial \bm{C}_e}\bm{F}_e^T -2q\bm{1}\\
    &=2\rho\ \mathrm{dev}\left(\bm{F}_e\frac{\partial \psi_r}{\partial \bm{C}_e}\bm{F}_e^T\right) +\left(\frac{2\rho}{3}\mathrm{tr}\left(\bm{F}_e\frac{\partial \psi_r}{\partial \bm{C}_e}\bm{F}_e^T\right)-2q\right)\bm{1}.
\end{aligned}
\end{equation}

By introducing $p = -\frac{2\rho}{3}\mathrm{tr}\left(\bm{F}_e\frac{\partial \psi_r}{\partial \bm{C}_e}\bm{F}_e^T\right)+2q$, we have the constitutive relation for $\bm{T}$:
\begin{equation}
    \bm{T} = 2\rho\ \mathrm{dev}\left(\bm{F}_e\frac{\partial \psi_r}{\partial \bm{C}_e}\bm{F}_e^T\right) -p\bm{1}.
\end{equation}
From this expression, it is evident that $p = -\frac{1}{3} \mathrm{tr}(\bm{T})$ is the pressure field. Since $p$ depends on the arbitrary field $q$, it cannot be determined directly from the constitutive relation and must be obtained through appropriate boundary conditions.

\section{Thermodynamically consistency of the evolution of transformation deformation}
\label{app:flow_thermo}
\noindent In this section we will show that the evolution relation of the transformation deformation given by \eqref{flow_rule} satisfies the thermodynamical consistent constraint \eqref{cons_flow}. Specifically, we show that the stress power due to transformation deformation $\bm{M}:\bm{D}_*$ is awlays larger or equal to zero.

According to the evolution relation of $\bm{D}_*$ in \eqref{flow_rule} and the definition of the Mandel stress in \eqref{mandel}, we can expand the stress power as following:
\begin{equation}
\label{ap2_1}
\begin{aligned}
    \bm{M}:\bm{D}_* &= \frac{\mu_H\dot{\alpha}}{2\mu^2}\left(\bm{F}_e^T\bm{T}_0\bm{F}_e^{-T}\right):\left(\bm{F}_e^{-1}\bm{T}_0\bm{F}_e^{-T} -\frac{\mathrm{tr}(\bm{F}_e^{-1}\bm{T}_0\bm{F}_e^{-T})}{\mathrm{tr}(\bm{F}_e^{-1}\bm{F}_e^{-T})}\cdot\bm{F}_e^{-1}\bm{F}_e^{-T}\right). \\
\end{aligned}
\end{equation}
Now based on the constitutive relation \eqref{const}$^1$, we can further simplify the first term in \eqref{ap2_1} as following:
\begin{equation}
\label{term1}
\begin{aligned}
  &  \left(\bm{F}_e^T\bm{T}_0\bm{F}_e^{-T}\right):\left(\bm{F}_e^{-1}\bm{T}_0\bm{F}_e^{-T}\right) \\
= & \bm{T}_0:(\bm{T}_0\bm{F}_e^{-T}\bm{F}_e^{-1})\\
=&\mu\bm{T_0} :\left(\bm{1}- \frac{1}{3}\mathrm{tr}(\bm{F}_e\bm{F}_e^T)\bm{F}_e^{-T}\bm{F}_e^{-1}\right)\\
=& -\frac{1}{3}\mu\cdot\mathrm{tr}(\bm{F}_e\bm{F}_e^T)\left(\bm{T}_0:\bm{F}_e^{-T}\bm{F}_e^{-1}\right)\\
=& -\frac{1}{3}\mu\cdot\mathrm{tr}(\bm{F}_e\bm{F}_e^T)\cdot\mathrm{tr}(\bm{F}_e^{-1}\bm{T}_0\bm{F}_e^{-T}),
\end{aligned} 
\end{equation}
and the second term in \eqref{ap2_1}:
\begin{equation}
\label{term2}
\begin{aligned}
   \frac{\mathrm{tr}(\bm{F}_e^{-1}\bm{T}_0\bm{F}_e^{-T})}{\mathrm{tr}(\bm{F}_e^{-1}\bm{F}_e^{-T})}\cdot\left(\bm{F}_e^T\bm{T}_0\bm{F}_e^{-T}\right):\left(\bm{F}_e^{-1}\bm{F}_e^{-T}\right) 
=  \frac{\mathrm{tr}(\bm{F}_e^{-1}\bm{T}_0\bm{F}_e^{-T})^2}{\mathrm{tr}(\bm{F}_e^{-1}\bm{F}_e^{-T})}.
\end{aligned} 
\end{equation}
Now substitute \eqref{term1} and \eqref{term2} back into \eqref{ap2_1}, we get:
\begin{equation}
\label{temp}
\begin{aligned}
    \delta = -\frac{\mu_H\dot\alpha}{2\mu^2}\mathrm{tr}(\bm{F}_e^{-1}\bm{T}_0\bm{F}_e^{-T})\left(\frac{\mu}{3}\mathrm{tr}(\bm{F}_e\bm{F}_e^T)+\frac{\mathrm{tr}(\bm{F}_e^{-1}\bm{T}_0\bm{F}_e^{-T})}{\mathrm{tr}(\bm{F}_e^{-1}\bm{F}_e^{-T})}\right).
\end{aligned}
\end{equation}
For the term $\mathrm{tr}(\bm{F}_e^{-1}\bm{T}_0\bm{F}_e^{-T})$, according to the constitutive relation \eqref{const}$^1$, we have:
\begin{equation}
\label{myeq}
    \begin{aligned}
        \mathrm{tr}(\bm{F}_e^{-1}\bm{T}_0\bm{F}_e^{-T}) = \mu(3-\frac{1}{3}\mathrm{tr}(\bm{F}_e\bm{F}_e^T)\cdot\mathrm{tr}(\bm{F}_e^{-1}\bm{F}_e^{-T})).
    \end{aligned}
\end{equation}
Now by substituting \eqref{myeq} into the second term in \eqref{temp} we have:
\begin{equation}
\bm{M}:\bm{D}_*=-\frac{3\mu_H\dot\alpha}{2\mu}
\frac{\mathrm{tr}(\bm{F}_e^{-1}\bm{T}_0\bm{F}_e^{-T})}{\mathrm{tr}(\bm{F}_e^{-1}\bm{F}_e^{-T})}.
\end{equation}

Note that both $\bm{B}_e = \bm{F}_e\bm{F}_e^T$ and $\bm{C}_e^{-1} = \bm{F}_e^{-1}\bm{F}_e^{-T}$ are positive-definite matrix and we have $\mathrm{tr}(\bm{F}_e^{-1}\bm{F}_e^{-T})>0$. Hence, the following inequality holds for any positive-definite matrix $\bm{A} (3\times 3)$:
\begin{equation}
    \mathrm{tr}(\bm{A})\cdot\mathrm{tr}(\bm{A}^{-1})-9\geq 0,
\end{equation}
and, according to \eqref{myeq}, we have:
\begin{equation}
    \mathrm{tr}(\bm{F}_e^{-1}\bm{T}_0\bm{F}_e^{-T}) \leq0.
\end{equation}

As a result, the following inequality always hold:
\begin{equation}
    \bm{M}:\bm{D}_*\geq 0,
\end{equation}
thus the evolution relation of the transformation deformation we proposed in \eqref{flow_rule} always satisfies the thermodynamical constraint \eqref{cons_flow}.

\section{Evolution of transformation deformation for a general choice of $\psi_r$}
\label{app:flow_general}

In this section, to simplify the expressions, we introduce $\tilde{\psi}_r = \rho \psi_r$, which represents the free energy per unit volume in the deformed configuration. The constitutive relation \eqref{cons_stress} gives the deviatoric part of the Cauchy stress $\bm{T}_0$:

\begin{equation}
    \bm{T}_0 = \mathrm{dev} (\bm{T}) = 2 \mathrm{dev} \left( \bm{F}_e \frac{\partial \tilde{\psi}_r}{\partial \bm{C}_e} \bm{F}_e^T \right).
\end{equation}

As shown in Fig. \ref{fig:kinematics}, the emergence of the transformation deformation is to match the mismatch of the newly formed polymer chains with the existing network. Considering a process where there is no change in macroscopic distortion, i.e., $\mathrm{dev}(\bm{L}) = 0$, we obtain the following constraint between the evolution of elastic deformation and transformation deformation:
\begin{equation}
    \dot{\bm{F}}_e = - \bm{F}_e \bm{D}_*.
\end{equation}

The flow rule states that, under this condition (no macroscopic distortion), the deviatoric part of the stress $\bm{T}_0$ should also remain unchanged:
\begin{equation}
    \dot{\bm{T}}_0 = \mathrm{dev} \left( \bm{F}_e \left( \frac{\partial^2 \tilde{\psi}_r}{\partial \bm{C}_e \partial \alpha} \dot{\alpha} 
    - \bm{D}_* \frac{\partial \tilde{\psi}_r}{\partial \bm{C}_e} 
    - \frac{\partial \tilde{\psi}_r}{\partial \bm{C}_e} \bm{D}_* \right) \bm{F}_e^T \right) = \bm{0}.
\end{equation}

This leads to the following equation to determine $\bm{D}_*$:
\begin{equation}
    \bm{D}_* \frac{\partial \tilde{\psi}_r}{\partial \bm{C}_e} + \frac{\partial \tilde{\psi}_r}{\partial \bm{C}_e} \bm{D}_* = 
    \frac{\partial^2 \tilde{\psi}_r}{\partial \bm{C}_e \partial \alpha} \dot{\alpha} - q \bm{F}_e^{-1} \bm{F}_e^{-T},
\end{equation}
where $q$ is a scalar that can be determined through the incompressibility condition:
\begin{equation}
    \mathrm{tr}(\bm{D}_*) = 0.
\end{equation}

\section{Derivation of the heat equation}\label{app:heat}
\noindent In this section we will show the derivation of the heat equation \eqref{heat}. The heat equation describes the evolution of the temperature field, which can be derived from the energy balance equation \eqref{energy_balance}:
\begin{equation}\label{heat_1}
    \rho\dot{\varepsilon} -\bm{T}:\bm{L} + \mathrm{div}\ \bm{q} -\rho r =0.    
\end{equation}

The internal energy $\varepsilon$ and the Helmholtz free energy $\varphi$ is connected by \eqref{Helmholtz}, which imples:
\begin{equation}\label{heat_2}
    \dot{\varepsilon} = \dot{\varphi} +T \dot{S} +S\dot{T}.
\end{equation}

The mechanical power $\bm{T}:\bm{L}$ can also be related to the time derivative of the free energy $\dot{\varphi}$:
\begin{equation}\label{heat_3}
\begin{aligned}
    \bm{T}:\bm{L} & = \frac{1}{2}\bm{S}:\dot{\bm{C}}_e + \bm{M}:\bm{D}_* - J_i^{-1}p\cdot \dot{J}_i \\
    &=\rho\frac{\partial \psi_r}{\partial\bm{C}_e}:\dot{\bm{C}}_e +\rho \dot{\psi}_i - J_i^{-1}p\cdot\left(\frac{\partial J_i}{\partial T}\dot{T} +\frac{\partial J_i}{\partial \alpha}\dot{\alpha}\right)\\
    &  = \rho\frac{\partial \psi_r}{\partial\bm{C}_e}:\dot{\bm{C}}_e +\rho \dot{\psi}_i +\rho\frac{\partial \psi_r}{\partial T}\dot{T}+\rho S\dot{T} +\rho\mathcal{F}\dot{\alpha} + \rho\frac{\partial \psi_r}{\partial \alpha}\dot{\alpha}\\
    & = \rho \dot{\psi} + \rho S\dot{T} +\rho\mathcal{F}\dot{\alpha}.
\end{aligned}
\end{equation}
Substitute \eqref{heat_2} and \eqref{heat_3} into \eqref{heat_1}, we got:
\begin{equation}
     T\dot{S} = \mathcal{F}\dot{\alpha}-\frac{1}{\rho}\mathrm{div}\bm{q} +r.
\end{equation}
With the constitutive relation for the entropy in \eqref{const}$^2$, we have:
\begin{equation}
    \dot{S} = \frac{c}{T}\dot{T} -\frac{\beta}{\rho_0}\dot{p}.
\end{equation}
As a result, the heat equation writes:
\begin{equation}
     c\dot{T} = \mathcal{F}\dot{\alpha}-\frac{1}{\rho}\mathrm{div}\bm{q} +\frac{\beta T}{\rho_0}\dot{p} +r.    
\end{equation}
\section{Material parameters for bulk and frontal polymerization}
\label{app:values}
\noindent In this section we present the representative material parameters for both bulk polymerization and frontal polymerization. These parameters are used in the numerical simulations of \eqref{eq:reac_diffusion} introduced in Section~\ref{sec:regimes}, related to the result shown in Fig. \ref{fig:mode}.
\begin{table}[H]
    \centering
\begin{tabular}{lccc}
\hline
Parameter & Bulk polymerization (PDMS) & Frontal polymerization (DCPD) & Unit \\
\hline
$c$      & 1440 & 1568 & kJ/(m$^{3}$·K) \\
$\kappa$ & 0.1736 & 0.152 & W/(m·K) \\
$H$      & $1.05\times 10^5$ & $3.67\times 10^5$ & kJ/m$^{3}$ \\
$C$      & $2.135\times 10^7$ & $1.8\times 10^{15}$ & s$^{-1}$ \\
$E_a$    & 64.5 & 110.75 & kJ/mol \\
\hline
\end{tabular}
    \caption{Representative thermal and chemical kinetic material parameters for bulk polymerization \citep{sain2018thermo} and frontal polymerization \citep{li2024mechanical}}
    \label{tab:parameter}
\end{table}

\section{Relation between $\bar{Y}$ and the Zeldovich number}
\label{app:Zeldovich}
\noindent In this section, we show that the dimensionless parameter $\bar{Y}$ introduced in Section~\ref{sec:stability} reduces to twice the Zeldovich number in the absence of heat loss and mechanical loading. Under adiabatic, stress-free conditions, we have $h = 0$ and $\lambda_e = 1$. Based on the expression for the cutoff temperature $T_*$ in \eqref{eq:cut_off}, the driving force $\mathcal{F}$ in \eqref{eq:driving_1d} and the length scale $s_2(v)$ in \eqref{eq:s1s2}, we obtain
\begin{equation}
T_* = T_0, \qquad \mathcal{F} = H, \qquad
s_2(v) = \frac{\kappa}{c v}.
\end{equation}

According to the definitions of $\bar{Y}$ in \eqref{eq:Y} and \eqref{eq:Y_bar}, the expression of $\bar{Y}$ under adiabatic, stress-free conditions is
\begin{equation}\label{eq:z_1}
    \bar{Y} =  \frac{{H}}{c T_m} \cdot\frac{C\kappa}{cv^2} e^{-\frac{E_a}{R T_m}}.
\end{equation}

Using the expressions for $T_m$ and $v$ obtained in \eqref{eq:velocity}, we have
\begin{equation}\label{eq:z_2}
    T_m = T_0 + \frac{H}{c}, \qquad v = \sqrt{\frac{C\kappa}{c} \int_{T_0}^{T_m} e^{-E_a / RT} \frac{\mathrm{d}T}{T}}.
\end{equation}

Substituting \eqref{eq:z_2} into \eqref{eq:z_1}, we obtain
\begin{equation} \label{eq:z_3}
    \bar{Y} = \frac{T_m-T_0}{T_m}\left(\int_{T_0}^{T_m} e^{-E_a / RT} \frac{\mathrm{d}T}{T}\right)^{-1} e^{-\frac{E_a}{RT_m}}.
\end{equation}

Note that the integration in \eqref{eq:z_3} $I =\int_{T_0}^{T_m} e^{-E_a / RT} \frac{\mathrm{d}T}{T}$ can be expressed in terms of the exponential integral, defined by
\begin{equation}
    E_1(x) = \int_x^{+\infty} \frac{e^{-u}}{u} du.
\end{equation}
Based on the substitution $u = E_a/(RT)$, we obtain
\begin{equation}
    I =\int_{T_0}^{T_m} e^{-E_a / RT} \frac{\mathrm{d}T}{T} = \int_{E_a/(RT_m)}^{E_a/(RT_0)} \frac{e^{-u}}{u}du = E_1\left(\frac{E_a}{R T_m}\right) - E_1\left(\frac{E_a}{R T_0}\right).
\end{equation}

For most cases of frontal polymerization, the activation energy $E_a$ is typically large.  The exponential integral admits the well-known asymptotic expansion  $E_1(x) \sim \frac{e^{-x}}{x}$ for large $x$. As a result, we have the following approximation for $I$:
\begin{equation}\label{eq:z_4}
    I \approx E_1(\frac{E_a}{RT_m}) \approx \frac{e^{-\frac{E_a}{RT_m}}}{E_a/(RT_m)}
\end{equation}

Substituting the approximation \eqref{eq:z_4} into \eqref{eq:z_3}, and using the definition of the Zeldovich number in \eqref{eq:zeldovich}, we obtain
\begin{equation}
    \bar{Y} = \frac{T_m-T_0}{T_m}\frac{E_a}{RT_m} = Z
\end{equation}

\bibliography{main}
\bibliographystyle{elsarticle-harv}
\end{document}